\documentclass{aastex}
\input{psfig.sty}

\newcommand{\etal}{et al. }

\def\koutara{{ \Huge \raise-0.5ex\hbox{$\Box$}}}
\def\sol{$_{\odot}$~}
\def\degr{\hbox{$^\circ$}}
\def\degree {$^{\circ}$}
\begin{document}

\title{ISO--LWS observations of the two nearby spiral galaxies: NGC6946 and NGC1313}
\tiny
\normalsize
 
\author {Alessandra Contursi\altaffilmark{1,2}, 
Michael J. Kaufman\altaffilmark{3,4}, 
George Helou\altaffilmark{1,2},
David   J. Hollenbach\altaffilmark{4},
James Brauher\altaffilmark{1},
Gordon J. Stacey\altaffilmark{5},
Daniel A. Dale\altaffilmark{1,2},
Sangeeta Malhotra\altaffilmark{6}, 
Monica Rubio \altaffilmark{7}, 
Robert H. Rubin\altaffilmark{4}, 
Steven D. Lord\altaffilmark{1}}
\altaffiltext{1}{IPAC, California Institute of Technology 100-22, Pasadena, CA 91125}
\altaffiltext{2}{Downs Laboratory of Physics, California Institute of Technology, Pasadena, CA 91125}
\altaffiltext{3}{Dept. of Physics, San Jose State University, San Jose, CA
95192-0106}
\altaffiltext{4}{Space Sciences Division, MS 245-3, Nasa Ames Research Center,
Moffett field, CA 94035}
\altaffiltext{5}{Cornell University, Astronomy Dept. 220 Space Science Building,
Ithaca, NY 14853}
\altaffiltext{6}{Johns Hopkins University, Baltimore, MD 21218}
\altaffiltext{7}{Departamento de Astronom\'ia Universidad de Chile, Casilla
36--D, Snatiago, Chile}
 
\begin{abstract}
   We present the analysis of  ISO-Long Wavelength Spectrometer (LWS) observations of the two nearby 
late-type galaxies NGC1313 and NGC6946. Both galaxies have been fully mapped in the [CII] 
Far-Infrared (FIR) fine structure
line at 158 $\mu$m and some regions have been observed also in the [OI](63 $\mu$m) and [NII](122 $\mu$m) lines.
We use these observations to derive the physical properties of the atomic interstellar medium,
to establish how  they relate with other interstellar medium components (gas and dust), and 
how they vary with 
different galaxy  components like nucleus,
spiral arms and disk.  \\
The [CII] line is the main cooling line of the atomic medium. In NGC6946 and
NGC1313, its  emission
represents   0.8 $\%$ of the infrared emission.
Moreover, the [CII] emission can be spatially associated with three components: the nucleus, the
star forming regions in spiral arms and the diffuse  
galaxy disk. This last component contributes  $\lesssim$40 $\%$ in NGC6946 and $\sim$30 $\%$ in NGC1313
to the total emission.\\
We apply the PDR model by Kaufman \etal \cite{Kaufman} to derive   PDR physical parameters responsible for the  
neutral atomic gas emission ($G_0$,  $n$ and $T_s$). The results  do not significantly differ from what 
Malhotra \etal \cite{Malhotra} found by
modelling the integrated emission of a sample of 60 normal galaxies. This 
suggests  that the emission  in each region under the LWS beam in NGC6946 and
NGC1313 (corresponding to
a linear size of $\sim$1.5 kpc)    is likely to arise from 
a mixture of  components similar to the mixture producing   
the  integrated emission of normal galaxies. However,   some  regions in NGC6946 have a $G_0$/$n$ ratio
   $\sim$2--3 times smaller than the mean value found for the normal
 galaxy sample (1.3), suggesting that the beam averaged contribution   of a less active component 
 in these regions  is higher than its contribution in the integrated emission
 of normal galaxies or, conversely, that the bulk of the integrated emission of the normal
 galaxies is dominated by a few active regions probably located in their nuclei.\\
 CO(1--0)  and [CII]   in NGC6946  are well correlated   and the mean [CII]/CO
 ratio agrees with the mean integrated ratios of the normal galaxies sample. This value ($\sim$ 500) is a
 factor $\sim$2 less than the mean ratio found for a  sample of   normal galaxies observed with KAO by Stacey
 \etal \cite{Stacey}. This difference is probably due to the fact that  
 the  KAO beam  (55$\arcsec$) is smaller than the LWS beam (75$\arcsec$),
such that the  Stacey \etal \cite{Stacey}  KAO observations
  are likely to be more   biased towards the nucleus of the galaxies and
 therefore towards more active regions. In NGC1313 only 4 LWS regions have been observed in
 CO(1--0), and three of them detected.
 The [CII]/CO(1--0) seems to systematically increase from the north-east to the south, along the S-shaped
 spiral arm, indicating that the interstellar medium conditions in NGC1313 are much more
 inhomogeneous than the conditions in NGC6946. \\
HI and [CII] in NGC6946 are completely de-correlated, probably because
they arise from different gas components: [CII] arises  principally in dense and warm PDR and
 HI from diffuse ($n\lesssim 3\times 10^3$ cm$^{-3}$) gas. On the other hand, in NGC1313
  we successfully detect two distinct gas components:  a 
 cirrus-like    component where HI and [CII] are weakly correlated as observed
 in our Galaxy,  and a component
 associated with   dense  PDRs  completely de-correlated form HI as observed in NGC6946.\\
 Finally, we find that the   
 HI   residing in dense gas surrounding the star forming regions and presumably
 recently photo-dissociated,
 constitutes a few percent of the total HI. In turn,
 this dense gas  component  produces  most of the [CII]   emission  emitted by
 the atomic neutral medium, 
 even if its contribution is  lower  in NGC1313   than in NGC6946. On the other hand, the 
  [CII] emission arising from ionized gas is higher in NGC1313 than
in NGC6946.

\end{abstract}

 
\setcounter{footnote}{0}
\large{
\section{ INTRODUCTION}

The evolution of a galaxy is primarily summarized in its star formation
history, namely the rate at which gas is converted into stars, and where
within the galaxy these stars form.  These two quantities depend on the
thermodynamical conditions and chemical composition of the star-forming
interstellar medium (ISM).  Stellar radiation and evolution will in turn
modify the ISM conditions, injecting electromagnetic and mechanical energy,
and newly processed material into the ISM.  The interplay between ISM and
star forming processes is therefore the key to understanding galaxy evolution.
One way to approach this topic is to study the energy budget in nearby
galaxies, {\it i.e.} the energy input provided by stars to the ISM and the
resultant cooling of the ISM.  Interstellar dust is heated by stellar
radiation and cools by re--radiating at infrared wavelengths.  The heating
and cooling of the interstellar gas are more complicated processes because,
they depend on its phase (molecular, atomic, neutral or ionized), chemical
composition and temperature.

In this paper we use the main atomic gas cooling lines to investigate
the energy budget between stars and gas and the physical conditions of the
neutral atomic gas in two late type galaxies: NGC6946 and NGC1313.  These
two galaxies belong to a wider Infrared Space Observatory (ISO)--Key
Project sample on 69 normal galaxies{\footnote{The galaxies of this sample
have been selected to cover a large range in B/FIR ratios, morphologies,
infrared luminosities (L$_{FIR}$$<$$10^{12}$ L\sol) and to have IRAS colors
characteristic of normal galaxies.}}  observed with various ISO instruments
(Dale \etal \cite{Dale00}, Lu \etal \cite{Lu}, Malhotra \etal
\cite{Malhotra}).  In particular, 60 galaxies of the sample were observed
in several far infrared (FIR) fine structure lines with the
Long--Wavelength--Spectrometer (LWS)  instrument on board ISO (Malhotra \etal
\cite{Malhotra}).  Since most of the galaxies in the sample are smaller
than the LWS beam (FHWM=75$\arcsec$) or of comparable size, most of these
ISO--LWS observations provide the integrated heating/cooling properties of
the interstellar atomic gas (Malhotra \etal \cite{Malhotra}).  However,
NGC6946 and NGC1313 are sufficiently close and extended that they can be
fully mapped in the strongest fine structure line: the [CII] at 158 $\mu$m.
NGC6946 was also mapped in the [OI] line at 63 $\mu$m in its central $\sim$
6$\arcmin$$\times$6$\arcmin$ while only 3 positions were observed at this
wavelength in NGC1313. The centers and some other regions in the spiral
arms were also observed in the [NII] and [OIII] lines at 122 $\mu$m and 88
$\mu$m.  The ISO--LWS observations of NGC6946 and NGC1313 aimed at 
investigating how the main ISM physical processes change depending on which
component of the galaxy (diffuse disk, spiral arms, bulge) and/or which
phase of the interstellar medium is involved.

[CII] and [OI] are   the main cooling lines of the neutral atomic gas. 
In normal galaxies, [CII]
usually dominates energetically, carrying a luminosity  $\sim$ 0.3--1 $\%$ of the 
FIR luminosity (Crawford \etal \cite{Crawford}, Stacey \etal 
\cite{Stacey}).  The [CII] line arises from the $^2$P$_{3/2}$ $\rightarrow$ $^2$P$_{1/2}$
magnetic dipole transition of C$^+$ at $\frac{{\Delta}E}{k}$=91~K above the
ground state. As the ionization potential of the C atom is 11.3 eV, it can
be easily ionized by far UV (FUV) photons which escape from HII regions
(h$\nu$$<$ 13.6 eV).  The [OI] $^3$P$_{1}$ $\rightarrow$ $^3$P$_{2}$ fine
structure line at 63 $\mu$m has an excitation energy
$\frac{{\Delta}E}{k}$=228 K.  The C$^+$ ions and the O$^0$ atoms are
excited by collisions with atomic hydrogen in the neutral atomic gas, with
molecules (mainly H$_2$) in the molecular gas and (for C$^+$) with
electrons in HII regions. The critical density is  the density at which collisional 
de--excitation balances radiative de--excitation (Osterbrock \cite{Osterbrock}).
The critcal density  for collision of C$^+$ and O$^0$ 
with H atoms are: n$_{crit}^{[CII]}$ = 3$\times$10$^3$ cm$^{-3}$ (Crawford \etal
\cite{Crawford}) and
n$_{crit}^{[OI]}$= 5$\times$10$^5$ cm$^{-3}$ (Table 4 of Tielens and Hollenbach
\cite{Tielens}).

[OI] and [CII] emission from the neutral ISM arise in 
photodissociation regions (PDRs), {\it i.e.} those regions in which the
chemical and heating processes are dominated or induced by interaction with
FUV photons.  In PDRs the gas is heated by energetic photo--electrons ejected
by dust grains after absorption of FUV (6 eV $<$ h$\nu$ $<$ 13.6 eV)
photons (Hollenbach $\&$ Tielens \cite{Hollenbach}) or, for dense
($n$$>$10$^4$ cm$^{-3}$) gas regions, via collisional deexcitation of
vibrationally excited H$_2$ (Sternberg $\&$ Dalgarno \cite{Sternberg}).
Since smaller grains are more efficient at photo--electron production, the
smallest grains (sometimes referred to as Very Small Grains) or even large
molecules (those responsible for the  Aromatic Features observed in  Emission
(AFEs)  between 5 $\mu$m  and 20 $\mu$m often attributed to
Polycyclic Aromatic Hydrocarbons or PAHs) are likely to be the main sources
of photo--electrons and thus of gas heating in PDRs (Bakes $\&$ Tielens
\cite{Bakes}, Helou \etal \cite{CAFE}). The brightest PDRs in the FIR fine structure lines and in
the infrared continuum are the dense and warm regions of interface between
star forming regions and their parental molecular clouds.  However, we also
expect [CII] and [OI] from atomic ISM ({\it e.g} atomic clouds) and [CII]
from diffuse ionized medium (DIM) and HII regions. It is still not clear
how much of [CII] arises from each gas phase in galaxies.  For example, in
the Galactic interior
Heiles \cite{Heiles} ranked the DIM to be the most significant
contributor to the observed [CII] emission in the atomic medium  followed by
the atomic clouds and by the regions of interface between star forming
regions and molecular clouds.  Madden \etal \cite{Madden} claimed that
$\sim$ 70$\%$ of the [CII] emission in NGC6946 comes from both diffuse
neutral (HI) and diffuse ionized ISM. On the other hand, Sauty, G\'erin $\&$
Casoli \cite{Sauty}, using radiative transfer models of the same galaxy,
claim that only 30$\%$ of the [CII] emission arises from these ISM phases.
 
Using the LWS observations of  NGC6946 and NGC1313 
we address three topics:\\
1) We investigate whether the [CII] emission from different regions inside NGC1313 
and NGC6946 behaves like the integrated [CII] emission of the galaxies belonging 
to the ISO--KP sample of normal galaxies presented by Malhotra \etal
\cite{Elliptical}, \cite{Malhotra}.
\\
2) We take   advantage of the relatively high spatial resolution to
evaluate the relative contribution of various  phases of  ISM
 to the [CII] emission.
\\
3) We derive the local interstellar radiation field and the gas density of the
neutral PDRs analyzed inside NGC6946 and NGC1313, applying the PDR model from Kaufman
\etal \cite{Kaufman}.

The paper is organized as follows: Sec. 2 presents the main characteristics
of NGC6946 and NGC1313. Sec. 3 reports on the data reduction and
analysis. Results are presented in Sec. 4, where first we study the
relation between gas cooling, dust emission and the interstellar radiation
field (ISRF) intensity and where we compare the cooling inside the observed
galaxies with the global cooling of normal galaxies. We tentatively
estimate the contribution to the gas cooling associated with nucleus,
spiral arms and diffuse disk.  The discussion is presented in Sec. 5 where
we derive the physical parameters of the PDRs in the two galaxies by
comparing the PDR models by Kaufman \etal \cite{Kaufman} to the
observations. These parameters allow us to discuss the origin of the
observed [CII], HI and CO emission in both galaxies.  Conclusions are
summarized in Sec. 6.

\section{ THE OBSERVED GALAXIES}

\subsection{ NGC1313}
NGC1313, 3.6 Mpc away, is  an Sbd galaxy (Table \ref{tab1}) with a  metallicity a factor four
less than that in the Milky Way{\footnote{ We assume $< 12 +
log({N(O)\over N(H)})>$ = 9.04 for the Galaxy as in Garnett $\&$ Shields
\cite{Garnett}}}   
($12+log({N(O)\over N(H)})=8.4$, Walsh $\&$ Roy \cite{Walsh}). Authors classify
NGC1313 between a late type spiral and a Magellanic Irregular.
It is the highest mass 
barred galaxy known with no radial abundance gradient (Moll\'a $\&$ Roy
\cite{Molla}). Its HI emission
extends 24 kpc, well beyond the stellar component and  it exhibits HI 
superbubbles typical of Magellanic--Irregular galaxies (Ryder \etal \cite{Ryder}). The MIR emission
 (7 and 15 $\mu$m) observed with ISOCAM  extends only for the central $\simeq$ 8 kpc region (Dale \etal
\cite{Dale99}) and 
its distribution and extension agree with those in the H$\alpha$ line ( Ryder
\etal \cite{Ryder})
and at 1600 $\AA$ ( Kuchinski \etal \cite{Kuchinski}, this image not  shown in
this paper was kindly provided by B. Madore). NGC1313 was observed in the $^{12}$CO(1--0) line by 
Bajaja \etal \cite{Bajaja}. They report only a 3$\sigma$  upper limit
in the central region of the galaxy. Recently, we re--observed this galaxy
in the $^{12}$CO(1--0) and $^{12}$CO(2--1) lines  with the
new more sensitive receivers at SEST (Rubio \etal in preparation). 
We successfully detected molecular gas
in the north-east spiral arm, in the center and $\sim$3$\arcmin$ 
 south--west  from the center. 
No CO was  detected  in the   south  region. 

An area of $\sim$ 10$\arcmin$$\times$10$\arcmin$ corresponding to the
HI extension of NGC1313 was fully mapped with LWS at 158 $\mu$m in the
[CII] fine structure line as part of the ISO--KP on Normal galaxies allocated
with NASA time (Helou \etal \cite{ISOKP}). Figure 1 shows the full LWS coverage at this
wavelength (all  circles) on a HI map of NGC1313. The diameter of the circles
correspond to 75$\arcsec$, the adopted LWS beam size at 158 $\mu$m.
Three different pointings (blue circles in Fig. 1) were  re-observed also
in the  [NII(122$\mu$m)] and [OI(63 $\mu$m)] fine structure lines.  One 
of them,  close
to the galaxy center (region 89 in Fig. 1),  has been observed also 
in the [OIII(88 $\mu$m)] fine structure line.

\subsection{NGC6946}
NGC6946 is a nearby low inclination (i$\sim$30\degree) 
Scd galaxy at a distance of 4.5 Mpc  (Table \ref{tab1}) with an 
abundance  gradient d[O/H]/dR= -0.089$\pm$0.003 dex/kpc very similar to the
gradient in the Milky Way equal to -0.07$\pm$-0.015 (Belley $\&$ Roy
\cite{Belley} and references therein).
It has a nuclear starburst and 
a weak central bar with an associated gas flow where star formation is 
taking place (see Elmegreen, Chromey $\&$ Santos \cite{Elmegreen} and
references therein).
This galaxy shows a well pronounced open spiral pattern as well as extended
emission visible in the HI line, radio continuum, FIR, and MIR emission.
 Its north--eastern arm is the brightest arm at all wavelengths indicating 
a site of vigorous star formation.
Previous KAO observations in the [CII] line were presented by 
Madden \etal \cite{Madden}. These data show a maximum of emission in the starburst
 nucleus, a component well correlated with the main spiral arms, and  a 
 smooth component which the authors associate with the   diffuse 
 medium (both neutral and fully ionized) and which accounts for 
$\sim$ 70$\%$ of the total [CII] emission. The $^{12}$CO(1--0) and
$^{12}$CO(2--1) emission
 from NGC6946 is rather extended over the galaxy and it is mostly 
associated with the nuclear region and the Spiral Arms. Casoli \etal
\cite{Casoli} found a low $^{12}$CO(2--1) over $^{12}$CO(1-0) ratio, 
averaged in 23\arcsec~ beam size, of about 0.4
in the NGC6946 disk, 
suggesting H$_2$ gas densities between 150--1500 cm$^{-3}$ associated
 with cloud envelopes rather than cloud cores. The same ratio is $\sim$ 1 for the 
central starburst region, which
points to higher density and excitation conditions in the nuclear starburst, as expected.
  The HI surface brightness variation is   also low (a factor of $\sim$ 2-3) in
a $\sim$ 25\arcsec~ beam, as
reported in Boulanger $\&$ Viallefond \cite{NGC6946HI} corresponding to a
minimum and maximum deprojected HI column density of 3.6$\times$10$^{20}$ and
9.5$\times$10$^{20}$ cm$^{-2}$ respectively. In contrast to the molecular gas, which
peaks on the central starburst, the HI distribution has a minimum in this
region. 
Finally, ROSAT X--ray observations of NGC6946 show no evidence of hot
diffuse interstellar gas (Scott \etal \cite{Scott}).
The bottom  panel of Fig. 1 shows the LWS coverage at different wavelengths
superposed on the HI map of NGC6946.

\section{ LWS  data reduction and analysis}
The data were processed through the LWS Pipeline (OLP) Version 7.0.  Most
of the regions observed in NGC1313 and NGC6946 are in the faint flux 
regime (F$_{60 \mu m}$$< $50Jy in the 75$\arcsec$ beam).  Therefore, 
the continuum fluxes in the LWS spectra are affected by the errors in the
 dark current which can be of the same order as the continuum flux. 
 To correct for this, the dark currents were re-estimated and removed using
 the LWS Interactive Analysis (LIA).  However, since the dark currents are only additive
 in nature, they do  not affect our line flux estimates. With LIA the data
 were then corrected for any instrumental responsivity variations and the flux 
calibrated to the LWS  calibration source Uranus.  Glitches due to cosmic rays
 were removed from the data using the ISO Spectral Analysis Package (ISAP).  
 For a complete description of the LWS observation modes see Brauher (in
 preparation).
Spectral scans were co-added and averaged together using a 3-$\sigma$ clip
 in spectral bins  of about 0.05 $\mu$m.  For the L01 observation of the center
 of NGC6946, a  sinusoidal fringe associated with the LWS instrument was
 removed using a  defringing algorithm supplied within ISAP.  All spectral
 lines in our  study are unresolved.  After fitting a linear baseline to 
the data, the line fluxes were calculated assuming a Gaussian line with an 
effective instrumental profile of 
0.29$\mu$m for $\lambda$$<$ 90$\mu$m and 0.60$\mu$m for
 $\lambda$ $>$ 90$\mu$m ($\Delta$$v$$\sim$10$^3$ km s$^{-1}$).
  The uncertainties associated with each line flux were estimated as the product of 
this  effective instrumental width and the {\it r.m.s} of the linear baseline fit
 to the data.  In the case of non-detections, 3-$\sigma$ upper limits are
 calculated. 

 The final  flux measurements with associated uncertainties of the FIR lines observed 
  are shown in Tables \ref{tab2} and \ref{tab3} for NGC1313 and Table \ref{tab4} for 
NGC6946. The uncertainties  listed  are those derived from the line and baseline fitting
only.
They do not include the absolute uncertainties   $\sim$30$\%$.\\
As shown in Fig. 1, 
 we observed the two galaxies in the [CII] line  over their optical extent
 (R$_{25}$). However,
since the LWS pointings do not satisfy 
the Nyquist sampling criterion  and the LWS beam at 158 $\mu$m is 
significantly  smaller than the diffraction value,
we will not use the interpolated maps for the quantitative analysis. We instead 
use the single pointing measurements. To compare the measured flux in each LWS
pointing with  the  emission at other wavelengths (21 cm, $^{12}$CO(1--0),HiRes IRAS, ISOCAM)
we proceed as  follows. 
We assume a Gaussian LWS beam whose FHWM is 75$\arcsec$, the  mean value
published in the latest LWS technical report (Gry, Swinyard, Harwood \etal
\cite{Gry}). We then extract   
 the flux at other wavelengths with two methods: 1) on the images smoothed to
the LWS resolution by multiplying the peak value by the assumed beam area
2) by summing the flux from the (unsmoothed) mid  and far infrared 
continuum, HI and CO line images using a Gaussian
weighting with a  $\sigma_{gauss}$ determined by the 158 $\mu$m beam characteristics,
centered at each 158 $\mu$m pointings. The two fluxes thus obtained are 
compared for each wavelength and show typical  differences of about
1$\%$.  However, these two methods differ by  7$\%$ in two cases, 
namely the HI map, and the IRAS High
resolution images. This is due to the fact
that the PSFs of these images are not  perfectly circular as assumed.

Another source of error arises from the  uncertainty on the adopted LWS beam size.
For each wavelength other than [CII], the map is convolved with the adopted
LWS beam and the resulting flux compared to the observed [CII] flux at each 
pointing.  In order to
estimate the uncertainty, we explore a range of LWS beam sizes corresponding
to $1\sigma$ on the adopted size, namely 70$\arcsec$, 75$\arcsec$, 80$\arcsec$ 
and 90$\arcsec$.  For each of these sizes $\Omega$, we recompute the fluxes and 
analyze the correlation with [CII], re-deriving the dispersion around the best 
fit of the form:

\begin{equation}
f(\lambda, \Omega) = \rm{a}(\Omega)~f([CII])^{\alpha}
\end{equation}

As $\Omega$ is varied, the dispersion changes too.  Assuming that the 
increase in this dispersion is primarily due to using the wrong value of 
$\Omega$, we can estimate that contribution for the $\pm 1\sigma$
range as the contribution to be added in quadrature for the change in 
dispersion to obtain.

We performed this calculation for the ISOCAM images in LW2 and LW3, and 
for the 4 IRAS bands, and obtained uncertainties ranging from 9$\%$ to 16$\%$. 
We   stress that this procedure is not intended to find the correct 
ISOLWS beam size, but rather to estimate the uncertainty 
on the calculated fluxes due to an imprecise knowledge of
the  ISOLWS beam size. We also point out that this is an overestimate
of that uncertainty, since other effects might contribute to the change in
the dispersion, such as an intrinsically non-unit value of $\alpha$,
or the intrinsic gradients in the brightness of the galaxies at various
wavelengths.\\
In conclusion, the total percent
uncertainty in the flux at other wavelengths is $\sqrt{7^2+unc^2(\Omega,
\lambda)+rms^2(\lambda)}$ ,
where the $rms(\lambda)$ is the original noise in each image at each $\lambda$
divided by the flux in each LWS pointing.
 
A reference position has been observed several times for each galaxy providing
the background value. However,  inspection of
2.5\degree$\times$2.5\degree IRAS
images at 100 and 60 $\mu$m for both galaxies 
revealed   extended Galactic infrared emission in the direction
of NGC1313 only. No such   structure in the Galactic emission 
is evident in the direction of NGC6946. 
Fig. 2 shows
the HI contours superposed on the IRAS 100 $\mu$m image for NGC1313. Also marked in the
north--west
part of the frame is the position of the reference point observed with LWS.
It is clear that this reference point underestimates the average Galactic foreground emission
in the NGC1313 direction. The  Galactic contribution
estimation at 158 $\mu$m in the direction of NGC1313 is described
in Appendix 1 and it is equal to 11$\pm$5$\times$10$^{-14}$ erg s$^{-1}$ cm$^{-2}$ beam$^{-1}$.
For NGC6946 the average of 3 observations of the reference position is
31$\pm$7.5$\times$10$^{-14}$ erg s$^{-1}$ cm$^{-2}$ beam$^{-1}$. Thus the foreground Milky Way
emission is greater but more uniform for NGC6946.
 
We calculated the total  {\it r.m.s.} on the data measurements in NGC1313 summing in
quadrature  the {\it r.m.s} resulting from the line fitting
(Tables \ref{tab2} and \ref{tab3}) and the uncertainty on the foreground contribution 
derived from the dispersion of the distribution of the IRAS 
100 $\mu$m flux measured around NGC1313 as described in Appendix 1 
(5$\times$10$^{-14}$ erg s$^{-1}$ cm$^{-2}$ beam$^{-1}$).
In the following analysis we will consider
only those points with a [CII] flux higher than 5 times this {\it r.m.s.} after
foreground subtraction. The total uncertainties on the [CII] values include
also the 30$\%$ calibration uncertainties.\\
The uncertainties for each LWS measurements in NGC6946 are 
the combination of the uncertainty on the line plus baseline  
fitting (Table \ref{tab4}) and the 30$\%$ calibration uncertainties. For this galaxy
also, we will
consider only those regions with [CII] flux higher than  5 times
the {\it r.sm.s.} in the data (Table \ref{tab2}) after foreground subtraction.

Finally, the foreground corrected [CII] value was multiplied by 0.6 to take 
 account the fact that the LWS calibration has been performed
 on  point like sources whereas we are observing extended sources
  (Gry, Swinyard, Harwood \etal \cite{Gry}).\\
  We {\it interpolated} the [CII] flux corrected 
for the foreground emission to produce maps at 158 $\mu$m for display purposes
only. The [CII]  contours
superposed on  the LW2 (5--8 $\mu$m) ISOCAM image for both galaxies are shown 
in Fig. 3 and briefly discussed in Sec. 4.2.

 \section{ RESULTS}

\subsection{ [CII] emission, dust and radiation field: comparison with the ISO--KP sample}

Figure 4  shows  the logarithm of the 
[CII]/FIR{\footnote{Here, 
the FIR flux is defined as in Helou \etal ({\cite{Helou88}): 
FIR (W m$^{-2}$)=1.26$\times$10$^{-14}$$\times$(2.58$\times$I$_{60 \mu m}$(Jy)+
I$_{100 \mu m}$(Jy))}}
and [CII]/$\nu$f$_{\nu}$(5--10$\mu$m)
ratios as a function of 60/100 $\mu$m IRAS colors for NGC6946 and NGC1313.
These ratios are compared to  
  the  integrated emission  of a sample of 60 normal galaxies
presented in Malhotra \etal \cite{Malhotra} and Helou \etal \cite{CAFE}.
 For NGC6946 and NGC1313, only
the 5$\sigma$ detections at 158 $\mu$m are considered. Furthermore, since the 
mid--infrared (MIR) size of  NGC1313 is smaller than 
the extent of the emission  at 158 $\mu$m, for this galaxy  
we   use only the LWS pointings included in the area detected at MIR wavelengths.

The ratios used in Fig. 4 have the following physical meanings.
Assuming, as is the case here, that the [CII] flux is larger than the [OI] flux,
the [CII]/FIR ratio is an indication of  the efficiency of grain photoelectric
heating. Grain photoelectric heating of the gas occurs when stellar photons
eject energetic  electrons from grains into the gas. The efficiency is defined
as the ratio of the heat energy delivered to the gas by the photo electrons to 
the  FUV photon  energy
delivered to the grains. [CII] and [OI] generally dominate the cooling
of gas in the regions where grains absorb the stellar photons. Therefore, if
[OI] is weak, the [CII]/FIR ratio approximates the efficiency of grain
photoelectric heating.\\
The 60/100 $\mu$m is indicative of the temperature of the grains emitting at FIR
wavelengths. If these grains are in thermal equilibrium{\footnote{This is true
only to first order because the emission from smaller
grains not in thermal equilibrium contributes to the FIR emission and it
is expected to be larger in the 60 $\mu$m band in relatively low radiation
field}}, their
temperature is set by the intensity of the radiation field. In this
condition,  the 60/100 $\mu$m traces the  intensity of the radiation  field.\\
The FIR emission is mostly due to relatively large ($\sim$0.01--0.1 $\mu$m)
dust grains, which absorb most of the stellar flux (see Table 2 from Dale \etal
\cite{SEDs}). However, the photoelectric heating efficiency of small grains is
considerably higher than that of the large grains, so that the small grains
disproportionately contribute to the photoelectric heating of the gas (Watson
\cite{Watson},  Bakes $\&$ Tielens
\cite{Bakes}, Weingartner $\&$ Draine \cite{Draine}). The small grains and the
carriers responsible for the Aromatic Features in Emission (AFEs) 
produce most of the emission in the
5-10 $\mu$m wavelength range. The carriers responsible for AFE  are believed to be planar aromatic
compounds mostly associated with Polycyclic Aromatic Hydrocarbons ("PAHs", Puget
$\&$ L\`eger \cite{Puget}, Allamandola, Tielens $\&$ Barker \cite{Allamandola}) with sizes
$\lesssim$10 \AA.  We calculate the flux in the 5-10 $\mu$m band,
$\nu$f$_{\nu}$(5--10$\mu$m), by multiplying the ISOCAM LW2 (5--8.5 $\mu$m) flux
by a scale factor of 1.7 (see Helou \etal \cite{PHOTS},
\cite{CAFE}).

Figure 4 shows that the [CII]/FIR ratio decreases as the radiation field 
increases 
(Malhotra \etal \cite{Malhotra}). On the other hand the 
[CII]/($\nu$f$_{\nu}$(5--10$\mu$m)) ratio is quite constant for a large
range of radiation fields (Helou \etal \cite{CAFE}) (stars  in Fig. 4). 
The behaviour of the CII/FIR ratio has been
 interpreted as principally due  to an increase of the positive charge
 of grains
 as the incident stellar flux  increases, which lowers the photoelectric yield 
and thus the heating (and the cooling) of the gas (Malhotra \etal
\cite{Malhotra}). 
Bakes $\&$ Tielens
\cite{Bakes}, (see their Fig. 6) showed that, for a given spectrum of incident
stellar flux, 
 disk shaped grains,  like the carriers responsible for the AFEs, 
are less ionized than spherical grains of the same size. Other than this shape
effect, grain
charge is fixed by the G$_0$$\sqrt{T}$/$n_{e}$ ratio (Hollenbach $\&$ Tielens
\cite{Hollenbach}). Here G$_0$ is the FUV (6--13.6 eV) ISRF 
normalized to  the solar neighbourhood value expressed in Habing flux:
1.6$\times$10$^{-3}$ erg s$^{-1}$ cm$^{-2}$;  n$_e$ is the electron density in the gas.
  The grain charge increases with the 
G$_0$$\sqrt{T}$/$n_{e}$ ratio.  G$_0$$\sqrt{T}$/$n_{e}$ is proportional to the ratio
 of the UV photoejected
rate of electrons from a grain surface to the recombination rate of electrons
from the gas. As G$_0$$\sqrt{T}$/$n_{e}$ increases and all grains become more
positively charged, the heating efficiency of the larger grains is reduced more
than that of the carriers responsible for the AFE and, therefore, such carriers 
provide an increasingly  important fraction of the heating source for the gas.

The constancy of the [CII]/$\nu$f$_{\nu}$(5--10$\mu$m) ratio as the 
radiation field increases may suggest that there exists a 
special connection between  the carriers responsible for AFE and the 
photoelectric heating. Nevertheless, the nature of this connection
  is not yet completely understood. The fact that   both the excitation of 
  the aromatic carriers   
and C$^+$ heating are related to FUV photons  is not sufficient to explain
the observed MIR-[CII] relation because   neither the [CII]  nor the AFE 
  are
proportional to the FUV photons: the first because of the grain charging, the
second because, since   photons determine the charge states of the aromatic
carriers, there would be   proportionality between  AFE and  FUV photons, 
only if all these states had   the same efficiency in converting   photons into
the AFE, which is unlikely. 
One additional complication is that the photo-electric efficiency of the 
aromatic carriers, as traced by the ([CII]+[OI])/AFE ratio, increases with the 
60/100 $\mu$m ratio (Helou \etal \cite{CAFE}),  suggesting an increse of the
photoelectric heating of AF carriers, rather than a decrease as  
 expected in high heating
intensity environments due to grains charging. However, this decrease 
may be also due to a decrease of the aromatic carriers with respect to the bigger 
grains, as seems to  suggest the observed decreasing of 
the AFE/FIR ratio with the 60/100 $\mu$m colour ratio (Helou \etal
\cite{Helou91}). The weak point of this   scenario is that one has to invoke 
feedback mechanisms able to make these decreasing rates exactly the same. Helou \etal
 (\cite{CAFE}) propose also an alternate scenario where a quiescent ISM component
 (the cirrus-like component as defined in  Helou \cite{Helou86}) dominates the
 observed heating/cooling at low 60/100 $\mu$m as expected, and an active component
 represented by the PDRs at the surface of the molecular clouds illuminated by
 young massive stars, gets over for high IRAS colors. This active component 
 has high G$_0$/n ratio and therefore less gas heating due
 to the grain charge, and  cools principally through [OI]. This would imply that
 in such ISM the aromatic carriers are depressed, as seems to be confirmed by
 the a  AFE/FIR ratio which is lower in dense molecular clouds than in the atomic
   medium (Boulanger \etal \cite{Boulanger96}).

 The [CII]/FIR observations of separate regions within  NGC6946 and NGC1313
as a function of 60/100 $\mu$m  
agree with  the general trend outlined  by the global [CII]/FIR ratios 
of the ISO--KP  galaxies. Unfortunately, since
 there are no (LWS beam averaged) regions inside the resolved galaxies 
with a 60/100 $\mu$m ratio 
 $\gtrsim$ 0.6, we cannot
check whether the [CII]/FIR deficiency at high 60/100 $\mu$m is still similar to the
global behavior observed in more in galaxies. However, small [CII]/FIR ratios have been already observed in 
high radiation field sources of the Milky Way by Stacey \etal \cite{Stacey}.
The global averaged values of [CII]/FIR for these two galaxies agree with the
ISO--KP galaxies average value. 
The global mean values and dispersion are:  Log$_{ISO--KP}$([CII]/FIR)=-2.42
$\sigma_{ISO-KP}$=0.18;  
Log$_{N1313}$([CII]/FIR)=-2.18 $\sigma_{NGC1313}$=0.25; 
and  Log$_{N6946}$([CII]/FIR)=-2.29 $\sigma_{NGC6946}$=0.17.

The [CII]/$\nu$f$_{\nu}$(5--10$\mu$m) ratio in the regions of NGC6946 and
NGC1313 also statistically agree with the mean global ratio of the ISO--KP
sample (-1.85$\pm$0.18). At the distances of these galaxies the LWS FWHM beam
corresponds to a linear length $\sim$1.5 kpc.
However, the mean  [CII]/$\nu$f$_{\nu}$(5--10$\mu$m) values for NGC1313
(-1.67$\pm$0.17) and NGC6946  and (-2.1$\pm$0.16)  differ by more than
one $\sigma$ in dex from each other.
 This is also shown in the top panels of Fig. 5, where the
distribution of the [CII]/FIR and [CII]/$\nu$f$_{\nu}$(5--10$\mu$m), for the two nearby
galaxies only, are presented.
In principle, a high  [CII]/$\nu$f$_{\nu}$(5--10$\mu$m) ratio may  be due to either
 a higher [CII] emission per HI atom or to a lower MIR
emission per HI atom  in NGC1313 than in NGC6946.  Fig. 5 also shows the surface
brightness distributions
of the FIR (left middle panel), $\nu$f$_{\nu}$(5--10$\mu$m) (right middle panel), 
[CII] (left bottom panel) and $\nu$f$_{\nu}$(5--10$\mu$m)/FIR ratio (right bottom panel)
for NGC6946 and NGC1313 only. 
The FIR flux distribution in NGC1313 is
comparable to, if not shifted to lower values than, 
the FIR flux distribution in NGC6946. 
On the other hand, the averaged AF flux distribution in NGC1313 is
significantly shifted to lower values than in
NGC6946, whereas the [CII] flux distributions are
comparable. The most significant difference,
however, is  the 
$\nu$f$_{\nu}$(5--10$\mu$m)/FIR ratio distribution, which is significantly
 lower in
NGC1313 than in NGC6946.
{\it This means that there is intrinsically less AFE in NGC1313
 than in NGC6946}, even though the emission of the relatively larger grains and the gas cooling 
flux are comparable in the two galaxies. We furthermore note that   the NGC1313 property  of having   
[CII]/$\nu$f$_{\nu}$(5--10$\mu$m) values higher than NGC6946 is also shared
by   the irregular galaxies of the ISO--KP sample (marked with a box in Fig. 4)  and by three regions in IC10 
(marked with asterisks in Fig. 4, Hunter \etal \cite{Hunter}). 
The fact that  these Irregulars may also have MIR surface brightness smaller than
in normal spirals is  suggested by the behavior of
the 7$\mu$m/15$\mu$m colors as a function of the MIR surface brightness of NGC1313, NGC6946
and IC10  presented in Dale
\etal \cite{Dale99}. While the shape of the three curves is the same, the 
surface brightness for both NGC1313 and IC10  are significantly lower than the 
MIR surface brightness reached by NGC6946. 

What causes the   AFE   to be  lower in NGC1313 than in NGC6946?
The infrared surface brightness is proportional to the 
product of the dust column density and the intensity (and hardness) of the
radiation field (Dale \etal \cite{Dale99}). But, since the 60/100 $\mu$m ratio 
traces the intensity of the radiation field,
and  regions with the same 60/100 $\mu$m ratio have different [CII]/$\nu$f$_{\nu}$(5--10$\mu$m)
ratios in NGC1313 and NGC6946, we must conclude that the   
column density of the carriers responsible for AFE in NGC1313 is lower than in NGC6946,
either as an intrinsic property of the dust, and/or as a result of greater radiation
field hardness   in NGC1313 than in NGC6946, leading to the destruction
of these carriers (Boulanger
\etal \cite{Boulanger}, Cesarsky \etal \cite{M17}, Contursi \etal \cite{N66}).\\
MIR low resolution spectra of
very metal poor dwarfs  show weak AFE over large portions of 
the galaxies (Madden \cite{Madden00}).  NCG1313 is not as extreme in
its metal content, as  the galaxies targeted  by Madden \cite{Madden00}, but  we may be seeing
the beginning of this phenomenon which becomes more and more important as 
the metallicity of the galaxies decreases. \\ 
Therefore, it is likely  that the lower metal content in NGC1313 compared to
NGC6946 accounts for  the  lower column density of the carriers producing the
AFE observed in NGC1313.
In low metallicity galaxies,    [CII] shells surrounding
CO cores are larger  than those in normal metallicity environments (Lequeux \etal \cite{Lequeux}).
This would translate in a I(CII)/I(CO) ratio higher in NGC1313 than in NGC6946.
As we will see later (Sec. 5.3) this is indeed the case for two of three regions
detected  in the $^{12}$CO(1--0) line in NGC1313 (Rubio et al. in
preparation).

As discussed above and in Helou \etal \cite{CAFE}, in PDRs
the carriers resposnible for AFE and small grains in general, are the most efficient
contributors to the photo-electric effect.  Since 
the [CII]/AFE ratios in NGC1313 are greater than the values in  NGC6946, NGC1313 must have 
a greater fraction of its  [CII] emission not produced by photelectric effect 
on grains. 
This extra [CII] emission is probably arising from HII
regions, as opposed to PDRs. In HII regions the gas 
is   primarily heated by  electrons from HII.
Following Molhotra \etal \cite{Malhotra} it is possible to estimate
the contribution of ionized gas  to the observed [CII] 
emission, scaling    the [NII] fine structure line measurements at 122 $\mu$m
by a factor which depends on the C/N ratio.
We perform the detailed calculation in $\S$ 5.1. Unfortunately, in the case of
NGC1313, only three regions have been observed in the [NII(122 $\mu$m)] line,
and therefore we cannot draw statistically significant conclusions. However,
in at least one of these regions,  the observed [CII] emission seems to arise
nearly completely from diffuse ionized gas, a condition   not even  reached
in the starburst nucleus of NGC6946.
In galaxies with high fractions of [CII] arising from HII regions, the [CII]/FIR
ratio does not reflect the grain photoelectric heating efficiency. Moreover, although the
[CII]/FIR distribution of the regions in NGC1313 and NGC6946 and the global 
ratios of the ISO--KP galaxies statistically agree, the average [CII]/FIR ratio
for NGC1313 is somewhat higher than for NGC6946 and, more generally, irregular
galaxies tend to have higher [CII]/FIR ratios than spirals. This suggests
that irregular galaxies with a greater star formation rate (SFR) per unit surface 
than spirals (e.g. the Magellanic Clouds: Rubio, Lequeux $\&$ Boulanger \cite{Rubio}), 
derive more of their [CII] emission from HII regions than 
spirals do.

\subsection{Global [CII] emission of NGC6946 and NGC1313}

\subsubsection{NGC6946}
The [CII] contours of the {\it interpolated} map of NGC6946 superposed on a LW2
(7$\mu$m) ISOCAM image (Dale \etal 2000a) are shown in Fig. 3 (top).
As mentioned before, we do not use the interpolated map for the analysis
but    it can be useful to derive the  general morphological
characteristics of NGC6946 in the 158 $\mu$m emission line.
As expected, the starburst nucleus is the brightest [CII] source, and a 
secondary peak is visible toward the north--east arms of NGC6946, which are 
known to be the brightest arms in H$\alpha$. A weaker enhancement in the [CII]
emission is also visible around the
bright MIR spots north--west from the center. Therefore, these secondary
peaks presumably
are related to star forming regions. In addition, an extended
and diffuse emission is also present throughout the disk; its extension  is
 comparable to
the extended emission found at MIR wavelengths with ISOCAM at  6.75 $\mu$m
and 15 $\mu$m.  A similar decomposition in three components was previously done 
by Madden \etal \cite{Madden} with   KAO observations of the same galaxy.

The total [CII] luminosity of NGC6946 is 3.7$\times$ 10$^7$ L$_{\sun}$. 
This value has been obtained by summing all the 5$\sigma$ detections
of NGC6946 at 158 $\mu$m and its uncertainty is dominated by the 30$\%$ calibration uncertainty
of LWS.  This method of calculating the total [CII] luminosity, however,   is
likely to underestimate the total flux, because we may be missing some
flux in-between the pointings.  To estimate
how much flux we miss,  we  proceed in different ways.
A first crude estimation can be  done  by multiplying  the flux by 
the ratio of the area not covered by
the LWS beam to the area covered by the LWS beam  among adjacent pointings. 
This gives 17$\%$ more flux than the above value.\\
For a more precise estimation, we use the linear correlation  between the MIR and the [CII] emission presented in
Helou \etal \cite{CAFE} and introduced in the previous Section.\\
The mean log([CII]/($\nu$ f$_{\nu}$(5--10 $\mu$m))) value for the ISO--KP galaxies 
 is equal to -1.85 with a dispersion equal
to 0.18 dex.  Assuming a total flux in the LW2 filter equal to 11.19$\pm$2.24 Jy
(Dale \etal \cite{Dale00}) and using this mean ratio to estimate the [CII] flux,
 we obtain a total [CII] luminosity of 5.1$\times$10$^7$ L$_{\sun}$
with 20$\%$  uncertainty, {\it i.e.} the ISOCAM integrated flux uncertainty. Comparing 
this value to the value found considering
only the 5 $\sigma$ detections in the LWS pointings,
we conclude that we miss $\sim$ 30$\%$ of the total flux. Hereafter, we
will assume that the total luminosity of NGC6946 in the [CII] line emission is
5.1$\times$10$^7$ L$_{\sun}$ with a total uncertainty equal to 20$\%$. This
value compares favourably with the 5.2$\times$10$^7$L$\sun$ \footnote{The
luminosity 
values published in other papers  are calculated assuming a distance for NGC6946 
different  from 4.5 Mpc. Therefore, we give here the corresponding values 
scaled for the distance of 4.5 Mpc, 
assumed in this paper.} obtained in a similar
sized map from KAO observations (Madden \etal \cite{Madden}). 
This value corresponds to  0.8 $\%$ of the FIR luminosity. To calculate the ratio of the [CII] luminosity to
 the total infrared luminosity (TIR) we follow  Dale \etal \cite{SEDs} who calculated  TIR/FIR
as a function of the 60/100 $\mu$m IRAS ratios
by modelling the infrared emission of 69 normal galaxies observed by ISO and IRAS.
The   NGC6946  global 60/100 $\mu$m ratio is 0.46 and the corresponding TIR/FIR ratio is 2.26.
Thus the [CII] luminosity is  0.35$\%$ of the total infrared
luminosity.
   
Madden \etal \cite{Madden} decomposed the KAO map  of NGC6946 
 in three main [CII]  components:  the nuclear region,
the spiral arms, and an extended component (hereafter N, SA and E). We obtain a
nuclear [CII] luminosity (within
75 $\arcsec$) equal to 4$\times$10$^6$ L$_{\sun}$, 8 $\%$ of the total
flux.
Madden \etal \cite{Madden} found a luminosity equal to 2.9$\times$10$^6$ L$_{\sun}$ in
the central 60$\arcsec$, 
which is 6$\%$ of the total [CII] luminosity measured with KAO. However, these two values agree within the LWS and
KAO uncertainties.
In fact, we expect the nuclear value calculated from the KAO data  to be somewhat
smaller because the authors  integrated the [CII] flux over  a solid angle $\sim$45$\%$  smaller
 than the LWS beam.

Since we cannot use the interpolated map for our analysis, we 
cannot easily separate the [CII] emission associated  with the spiral arms and the
extended emission. Madden \etal \cite{Madden} compared   the KAO [CII], 
FIR, CO and HI emission along a east-west cut of NGC6946. They found  
the [CII] emission in the disk of NGC6946 a factor $\sim$ 2 higher than 
in the nuclear region when scaled to the 
FIR and CO emission.  They interpreted this as due to an extended [CII]
component which accounts for 70$\%$ of the total emission.
In Fig. 6 we show a similar comparison.   Since we cannot use the interpolated map to
calculate the emission along one given direction, in order to include  as many LWS
pointings as possible, we  perform two cuts with P.A. equal to
45$\degr$ and 135$\degr$ respectively (see Fig. 1). We do not find any significant
excess of the [CII] emission as in  Madden \etal \cite{Madden}.
However, we   have fewer points than  Madden \etal \cite{Madden} due
to the larger LWS beam.\\
Therefore, we try to estimate more precisely the amount of diffuse [CII] emission 
as follows. 
 Madden \etal \cite{Madden} found that the [CII] surface brightness
for the extended emission in NGC6946 ranged from 1 to 2 $\times$10$^{-5}$
erg s$^{-1}$ cm$^{-2}$ sr$^{-1}$.  The minimum and maximum [CII] surface 
brightness detected  with
LWS and corrected for the foreground emission are 3.6$\times$10$^{-7}$ (reg. 7
in Table \ref{tab4}; not a 5 $\sigma$ detection) and  4$\times$10$^{-5}$ erg s$^{-1}$ cm$^{-2}$ sr$^{-1}$
respectively (reg. 31 in Table \ref{tab4}).  
The surface brightness values for the extended emission given by Madden
\etal \cite{Madden},
are $\sim$30 times higher than the minimum surface brightness detected with LWS
at 158 $\mu$m. All  but two (one of which is the nucleus) of the LWS  5 $\sigma$
 measurements, have surface brightness
lower than 2$\times$10$^{-5}$ erg s$^{-1}$ cm$^{-2}$ sr$^{-1}$. This suggests that
this value is too high to arise principally in the diffuse low density gas. 
However, if we consider   
1$\times$10$^{-5}$ erg s$^{-1}$ cm$^{-2}$ sr$^{-1}$ as maximum emission for the diffuse
component, we find that it contributes   $\sim$ 35$\%$ of the total [CII]
emission.

We can estimate the minimum contribution of the [CII] extended component to the
total [CII] emission in NGC6946 by  simply  assuming that the [CII] extended component is
uniformly distributed over the galaxy disk with a surface brightness
corresponding to the minimum [CII] flux we have detected ($>$ 5 $\sigma$)
 in NGC6946. This value, corrected for the foreground emission,  
is 1.3$\times$10$^{-6}$ erg s$^{-1}$ cm$^{-2}$ sr$^{-1}$ and corresponds to region 27
 of Table \ref{tab2}. Integrating this value over   the optical size of NGC6946 (see
Table \ref{tab1}),
 we obtain
a total diffuse [CII] minimum contribution equal to  7.6$\times$10$^6$ L$\sun$,
15$\%$ of the total [CII] luminosity.
Using the [CII]/($\nu$ f$_{\nu}$(5--10 $\mu$m)) relation we find that this  minimum [CII]
  surface brightness corresponds to a LW2 surface brightness equal to 0.17
MJy sr$^{-1}$. Note that the typical  surface brightness of
quiescent regions of the Milky Way like the high galactic clouds, with an average HI
column density of 2--3 $\times$10$^{20}$ cm$^{-2}$     is $\sim$ 0.04 
MJy sr$^{-1}$ (Dale \etal \cite{Dale00}).\\
It is more likely that the [CII] diffuse component is not uniform and that 
it might decrease
with  galactocentric distance. Following this line of thought, we can set an upper
limit to the diffuse [CII] component using the [CII]/($\nu$ f${\nu}$(5--10 $\mu$m)) 
correlation. We divide the ISOCAM LW2 image in concentric rings of 15$\arcsec$ width,
{\it i.e} $\sim$ twice the ISOCAM resolution at 7 $\mu$m,   
 and assume that the [CII] diffuse component in each ring is uniform with a
[CII] surface brightness corresponding to the minimum LW2 surface brightness 
observed  in each ring. Summing up all these components we find that the maximum  [CII]
 extended contribution is equal to 1.8$\times$10$^7$ L$\sun$ which is
37 $\%$ of the total [CII] flux. 

In conclusion, we found that the  diffuse [CII] emission contribution
to the total flux is much smaller ($\lesssim$ 40$\%$) than the value
found by  Madden \etal
\cite{Madden}. We note that our value 
agrees with that
found by  Sauty Gerin $\&$ Casoli \cite{Sauty} who, through radiative transfer
modelling of NGC6946, estimate  the diffuse extended component at  30$\%$ of the
 total [CII] emission. The discrepancy between this and Madden \etal \cite{Madden}  work may
arise from the fact that  the background emission  in the NGC6946 direction 
is much better determined with the LWS observations than with the KAO data, for which no specific 
observations of reference positions were performed. 
This may explain why the KAO
data are systematically higher than the LWS measurements.
In fact, if we assume as typical diffuse [CII] surface brightness the
upper limit value given in Madden \etal \cite{Madden} 
equal to 2$\times$10$^{-5}$ erg s$^{-1}$  cm$^{-2}$ sr$^{-1}$, 
and we assume the
diffuse component to be uniformly distributed over the optical size of NGC6946,
 we find a
diffuse contribution to the total emission $\sim$2$\times$35$\%$, which is
close to what these authors found.

\subsubsection{NGC1313}
Fig. 3 (bottom) shows the [CII] contours of NGC1313 superposed on a 6.75 $\mu$m ISOCAM
image (Dale \etal \cite{Dale00}). As in the case of NGC6946 we use the interpolated [CII] map only for general
morphological investigation and for display purposes. This map has been
obtained considering only the 5$\sigma$ detections where the $\sigma$ associated
with each pointing has been calculated   as described in Sec. 3. There is bright
[CII] emission
associated with bright infrared emission,  corresponding to the center,
the bar and the HII regions of the galaxy. These regions are visible as an S-shaped 
morphology in the figure. Three peaks are visible, two in the northern
 and one in the southern arm. There is also  diffuse emission 
which smoothly decreases  towards the outer disk.\\
The total [CII] luminosity is 9$\times$10$^6$
L$_{\sun}$, a factor 5 less than the NGC6946 luminosity. As in the case of  NGC6946, this
value has been obtained summing all the 5$\sigma$ detections. If we account
for the missing flux by correcting for the area
in-between the beam, we obtain a luminosity equal to 1.05$\times$10$^7$ L$\sun$, 
15$\%$ more than the previous value.
Unfortunately
we cannot use the [CII]/$\nu$f$_{\nu}$(5--10 $\mu$m) ratio to better estimate
 the true total 
[CII] luminosity of NGC1313 because the MIR extension of this galaxy is  
smaller than the size of the [CII] emission (see Fig. 1 top panel). Therefore,
we will take our value for the total [CII] luminosity as a lower limit. 
The
L$_{[CII]}$/L$_{FIR}$ ratio is $>$ 0.5$\%$, comparable with what we obtain for
NGC6946. Also the IRAS 60/100 $\mu$m colors for the two galaxies are exactly the
same, giving the same TIR/FIR ratio and thus a   value of 0.2$\%$ for the
L$_{[CII]}$/L$_{TIR}$ ratio.

NGC1313 does not have classic spiral arms or  a classic  nuclear region and therefore it
is very difficult to separate the [CII] contribution from different galaxy
components even using other observations than the 158 $\mu$m measurements to identify these
components. Its 6.75 $\mu$m ISOCAM image 
 does  show a diffuse, low surface brightness  emission, which ranges
from 0.06 to 0.1 MJy/sr. These values are comparable to the typical
surface brightness at 6.75 $\mu$m of cirrus in our and external galaxies (Dale
\etal \cite{Dale00}). The corresponding [CII] surface brightness 
calculated through the [CII]/$\nu$f$_{\nu}$(5--10 $\mu$m) relation corresponds to 7$\times$10$^{-7}$ --
1$\times$10$^{-6}$ erg s$^{-1}$ cm$^{-2}$ sr$^{-1}$. We note that the [CII]
 lowest surface brightness 
(foreground subtracted) detected with LWS (even if this is not a 5$\sigma$ detection)
is   7.8 $\times$10$^{-7}$ erg s$^{-1}$ cm$^{-2}$ sr$^{-1}$, and it falls in this range.
Thus if we assume that a uniformly diffuse low surface brightness emission
between 7$\times$10$^{-7}$ --
1$\times$10$^{-6}$ erg s$^{-1}$ cm$^{-2}$ sr$^{-1}$ is present   over the
galaxy's entire  optical
 disk, this would contribute  20--30 $\%$ to the total [CII] emission.

\section{Discussion}
The [CII] 158$\mu$m, [OI] 63$\mu$m and FIR continuum observations can be used 
to constrain the density, far ultraviolet (FUV) flux, temperature, thermal pressure 
and other properties of the neutral interstellar gas in galaxies (e.g. Kaufman \etal 1999). 
In order to analyze the emission from neutral gas, we first have to remove the 
contribution of ionized gas to [CII] emission, which is done by scaling the [CII] emission
from HII regions with the [NII] 122$\mu$m emission from these regions (\S5.1). 
In \S5.2, we apply PDR models to the emission from neutral media. In \S5.3, we discuss the
relationship of [CII]--emitting gas with the neutral gas responsible for CO and HI emission, and 
in \S5.4 we assess the relative contributions of dense and diffuse neutral gas to the observed
HI emission.     

\subsection{[CII] emission from neutral and ionized gas}
[CII] emission can arise both from neutral (PDR) gas and from ionized (HII)
gas. In order to compare our observations to the PDR models, we must first
attempt to remove the [CII] emitted from ionized gas. Following Malhotra \etal
\cite{Malhotra}, we can estimate the [CII] emission from ionized gas by comparing it
with [NII] 122$\mu$m emission. [NII] only arises from ionized gas since 
nitrogen has an ionization potential of 14.5 eV.  Given
the gas phase  abundance ratio of carbon to nitrogen, C/N, we can 
estimate the [CII] emission from HII regions by an appropriate scaling of 
the [NII] emission. Absorption line studies of diffuse interstellar 
gas in the Milky Way (Sofia \etal \cite{Sofia}, Meyer \etal \cite{Meyer}) find  
(C/N)$_{diffuse} =1.9$. In dense ionized regions, Rubin \etal
\cite{Rubin88},\cite{Rubin93} 
find an abundance  ratio (C/N)$_{dense}=3.8$. This abundance ratio has been 
shown to be  independent of metallicity in both normal and irregular galaxies 
(Garnett \etal \cite{Garnett99}), so we adopt the Milky Way diffuse and dense gas values for 
NGC6946 and NGC1313. With these abundance ratios, the [CII] intensity is 5.8
times the [NII] intensity if the emission is from the Diffuse Ionized Medium
(DIM); the [CII] intensity is 1.1 times the [NII] intensity if the emission
is from dense HII regions. In NGC6946, the observed [CII]/[NII] flux ratio
is $>7.7$ in all regions where both lines were observed. In NGC1313, there are
only upper limits on this ratio (Figure 11), $>4$ in one location and $>10$ in
the two others. If the [NII] emission arises only from dense ionized gas, then
very little of a galaxy's overall [CII] emission comes from ionized gas; if 
the [NII] emission comes only from diffuse gas, then ionized gas contributes
considerably to the integrated [CII] emission. Most ($\sim 70\%$) of the [NII]
122$\mu$m emission from the Milky Way comes from the DIM (Heiles \cite{Heiles}). 
If we  assume that 70\% of the of the [NII] emission from NGC6946 and NGC1313 comes
from diffuse ionized gas, then a conversion factor [CII]/[NII] 
= 0.7([CII]/[NII])$_{diffuse}$ + 0.3([CII]/[NII])$_{dense}$=4.3 is
appropriate. However, in the next section, we explore the range of PDR
conditions possible when assuming either that the [NII] emission is entirely
from diffuse ionized (DIM) or entirely from dense ionized (HII) gas. In 
NGC6946, 12 regions were observed in both [CII] 158$\mu$m and [NII] 122$\mu$m
emission, resulting in 11 [NII] detections and one upper limit. In NGC1313,
3 regions were observed in both lines, but only upper limits were found
for [NII]; for these regions, we correct the [CII] emission based on 
the upper limits, so our results are less certain than for NGC6946.   

\subsection{Comparison with the PDR models}

In order to further constrain the interstellar medium conditions in NGC6946 and NGC1313, 
we compare the observed infrared line and continuum emission with the PDR models of 
Kaufman et al. (1999). These models allow for the determination of the average 
gas density, $n$, the surface temperature of the emitting gas $T_s$, 
and the average far-ultraviolet (FUV) flux, $G_0$, illuminating the 
interstellar gas, where $G_0$ is measured in units of the Habing (1968) value for the 
average solar-neighbourhood FUV flux, $1.6\times 10^{-3}\,\rm erg\,cm^{-2}\,s^{-1}$.  The most commonly used diagnostics are the ratio
of intensities in the main neutral gas coolants, [CII] 158$\mu$m/[OI] 63$\mu$m, along with the 
line-to-FIR continuum ratio, ([CII] 158$\mu$m + [OI] 63$\mu$m)/FIR. The use of ratios eliminates 
the beam area filling factor as a parameter, assuming the [CII], [OI] and FIR emission come from 
the same clouds, and allows derivation of G$_0$ and $n$.  

The Kaufman et al. models were developed for comparison
with relatively nearby Galactic star forming regions, where PDR gas is illuminated from one side
by hot stars. However, these ISO observations involve beams which cover a galaxy region of size 
$\sim 1.5\,\rm kpc$. The observed
line emission from this large region arises from an ensemble of clouds with random orientations. As
a result, it is appropriate to modify the model results so as to approximate the emission
from clouds illuminated on all sides. The models show that [OI] 63$\mu$m is typically
optically thick, while [CII] 158$\mu$m and the FIR continuum are optically thin. As a 
result, a spherical cloud will emit [CII] and continuum radiation from both the side 
facing the observer and the far side of the cloud, while [OI] emission comes only from the
side facing the observer. We have modified the Kaufman et al. models to account for 
illumination from all sides {\footnote{We reduced by a factor of two the model
prediction or, inversely,   doubled the observed [OI] value. The total infrared 
emission has been calculated from the FIR as explained in Sec. 4.2.1 and 4.2.2.
However, the PDR model assumes that only photons with energy between  6 and 13  eV
 ionize grains leading to gas heating. On the other hand,  dust  heating
occurs by  FUV photons and by photons with h$\nu$ $<$ 6 eV.
In order to compare observations and model prediction   we have to divide the 
TIR by a factor of two, since the observed TIR includes heating by photons of
h$\nu$$<$ 6 eV, but models do not. 
Therefore, we apply the model prediction with 
([OI]$_{model}\times{\frac{1}{2}})/[CII]_{model}$ and
([OI]$_{model }$+ 2$\times$[CII]$_{model}$)/(2$\times$TIR$_{model}$) ratios.
}}, and we use these modified results to find the gas density 
and FUV flux in each of our pointings. Figure 7 shows the pointing and beam sizes in NGC6946. 
Since our observations are averages over the ISM
in each beam, the results for $n$ and $G_0$ that we derive are intensity weighted
averages over the beam. Comparing the observed [CII] intensity with that
predicted by the models for the derived $G_0$ and $n$ gives the area filling factor of 
the emitting regions,
$\Phi_A=I[CII]_{observed}/I[CII]_{predicted}$ (Wolfire et al. 1995). 

Since we are dealing with regions inside NGC6946 and NGC1313, we do not know {\it a
priori} which of the two components of the ionized gas dominates the [NII] emission. We expect 
different contributions depending on which component of the galaxy (nucleus, spiral
arms, diffuse disk) is located in our beam. We therefore derive values of $G_0$ and $n$
obtained upon removing the ionized gas contribution to [CII] emission assuming [CII] arises 
only from the DIM or only from dense HII regions. The results for both galaxies in these 
two limits are listed in Tables \ref{tab5} and \ref{tab6}. The percentage 
of [CII] emission arising in ionized gas (Col. 6) is comparable for regions 
in NGC6946 and NGC1313.
Only one region in NGC1313 (reg. 91) has a very high contribution to the
observed [CII] emission arising in ionized gas: for exemple, in this region,
when one considers the   DIM contribution, all the observed [CII] emission 
seems to arise in ionized medium.
It is worth mentioning that this same region of NGC1313, has not been detected
 in CO(1--0)  though its   [CII] emission is comparable to the emission of  reg.
  69, where the highest CO
content has been observed (\S5.3.1).
 
Figure 7 shows a comparison between the G$_0$ and $n$ values (Col. 2 and 3 in
Tables \ref{tab5} and \ref{tab6})
obtained in NGC6946 and NGC1313 with those obtained for the ISO--KP sample (Malhotra \etal 
\cite{Malhotra}). For NGC6946 and NGC1313,  we only show the case in which the [CII] observed 
fluxes  have been corrected for the contribution from DIM, the same correction applied 
to the ISO--KP normal galaxy sample. We also plot the point corresponding to G$_0$ and $n$ 
obtained upon averaging the [CII], [OI] and FIR emission over the NGC6946 disk.
The solutions show that conditions in regions inside NGC6946 are similar to those
in the ISO--KP galaxy sample, although none of the regions observed here match
the highest G$_0$, $n$ solutions in the ISO--KP sample. In the ISO-KP sample, all of the 
emission from a galaxy fit in an single  beam.  We might have expected the conditions in 
isolated regions of our two galaxies to be more extreme than the galaxy averages
from the ISO-KP sample. The LWS beam is, however,  large enough ($\sim$1.5 kpc at NGC6946 and NGC1313)   
that we are still averaging over many PDR components. The fact that our results for $G_0$ 
and $n$ are systematically lower than the largest values found in the ISO-KP sample may be telling 
us that emission from the distant sample galaxies is dominated by a few massive star forming regions,
perhaps from each galaxy nucleus. 

Comparing the values given in Tables \ref{tab5} for  NGC6946, we note that 
all but regions 33 and 45 have comparable gas volume densities but different
$\frac{G_0}{n}$ values, meaning that what principally drives  the differences 
inside this galaxy is a change in the  FUV flux illuminating the clouds  in the
 LWS beam rather than density variations.\\
 We   now to compare the PDR model results to the MIR emission properties in
 NGC6946. Figure 8 shows the regions for which we
obtained model solutions in NGC6946 overlayed on an ISOCAM image at 7 $\mu$m.
Most of the regions with the
 lowest ($<$0.2) $\frac{G_0}{n}$ values (15, 33, 40, 43 and 47), are also those
 not or partially related
 to bright spots at mid infrared (MIR) wavelengths.
  Thus it is likely that in these regions, the contribution of an active component to the total MIR
 emission is lower than in the rest of the galaxy indicating that, 
 to first order  bright MIR spots corresponds to warmer  PDRs as expected.
 However, since the MIR emission is proportional to the product of  dust
 (and therefore gas) column density and the intensity of the radiation field,
 its relation with the $\frac{G_0}{n}$ ratio is not straightforward.
 This is clearily  shown by the example
  of region 45, which has the highest $\frac{G_0}{n}$
 value but it does not correspond to any bright MIR emission.
 In this case the high $\frac{G_0}{n}$ value is due to a very low gas volume 
 density and a moderate G$_0$.
 On the contrary, bright MIR regions as region 47, can have a low
 $\frac{G_0}{n}$ value. In conclusion, to first order, MIR bright emission does
 correspond to warm and dense PDRs but this  is not a one to 
 one correspondence.
 \\
 The [CII] intensities from neutral gas in two modeled regions in NGC1313 are 
lower limits, since the contribution from diffuse ionized gas to the total [CII] emission has been scaled 
from the upper limits in the [NII] line.  Their $\frac{G_0}{n}$ values are comparable to
the lowest found in the disk of NGC6946. All these regions correspond to bright
MIR emission.

\subsection{Relationship between [CII], CO and HI emission}

[CII], CO and HI emission generally originate in PDRs. HI and [CII] are generated in 
diffuse clouds and from the surfaces of molecular clouds, with [CII] being
less sensitive
to the low density ($n\ll 10^3\,\rm cm^{-3}$) components; CO emission arises from deeper
into PDRs, where molecules are shielded from dissociating radiation but where FUV photons
still dominate the heating and/or dissociation of other oxygen--bearing molecules. 
In this section, we attempt to determine to what extent the [CII], HI and CO emitting
gases are physically related in our galaxies. 

Figure 6 already gives an indication of the origins of these lines in NGC6946. [CII], CO 
and the FIR all scale closely, peaking toward the center of the galaxy and falling off
considerably in the outer regions. HI 21 cm emission, however, remains quite constant in 
intensity over the mapped area. HI 21 cm emission is sensitive neither to density nor
temperature, but only to the column density of HI. [CII] emission increases sharply 
with density, to $n\sim n_{crit}\sim 3\times 10^3\,\rm cm^{-3}$ and with temperature,
to $T\sim \Delta E/k\sim 92~\rm K$. CO $J=1-0$ emission tends to be optically thick, and 
is therefore mainly dependent on the area filling factor of molecular clouds, with a rather
weak dependence on $G_0$ or  $n$ (Kaufman \etal 1999). The impression given by Figure 6 is
that since CO tracks the molecular gas, and since the molecular gas is where the star form that
power the FIR emission, the CO correlates with the FIR. The [CII] tracks this dense component,
as if the [CII] originates on the surfaces of the molecular clouds.  As we will see in \S5.3.1 below, there
are good theoretical reasons to expect this {\it spatial} correlation. 
 
Fig. 9 shows [CII] intensities versus $^{12}$CO(1-0)
intensities for regions in NGC6946 and NGC1313, and it compares these with 
the Stacey et al. \cite{Stacey} galaxy sample, as well as the ISO--KP galaxies for
which $^{12}$CO(1--0) data are available (Lord \etal in preparation).
The CO data for NGC6946 are from a $^{12}$CO(1-0) map kindly provided by M.D. Thornley, 
T. Helfer  and M. Regan  from the BIMA Survey on Normal Galaxies (SONG) 
sample. It covers about 6.5$\arcmin\times$6.5$\arcmin$ and the CO intensities
averaged over the LWS beam for those regions included in this area  are listed
in Table \ref{tab8}.
Also plotted are three regions of NGC1313 for which $^{12}$CO(1-0) data were taken
at SEST in November 2000. 
Since the SEST beam at 115 GHz is much smaller (40$\arcsec$) than the LWS beam at 158 $\mu$m,
we observed  three positions inside each  LWS beam (spaced 
at ${1}\over {2}$ the FHWM of the SEST beam) to try to estimate the
average surface brightness in the LWS beam
(Rubio \etal in preparation). 
These averaged intensities are listed in Table \ref{tab8}.

Fig. 10 shows the relationship between the deprojected\footnote{The inclination
angles assumed for NGC6946 and NGC1313 are 30$\degr$ and 48$\degr$ respectively}  [CII]   
surface brightness and the deprojected HI column density, calculated in the optically thin limit,
for each region in which we have a 5$\sigma$ [CII] detection. 
We also plot curves representing the intercloud medium (n$\sim$0.1
cm$^{-3}$ and T=10$^4$ K ), standard HI clouds (n$\sim$30 cm$^{-3}$ and
 T$\sim$100 K assuming an area filling factor equal to unity) and for 
cirrus emission in our galaxy (Bennett \etal \cite{Bennett}).  
The dependence on $n$ arises mainly because the HI is in {\it Local Thermodynamic Equilibrium} (LTE) 
even at the lowest interstellar densities, whereas the critical density for the [CII] transition is 
about 3$\times$10$^3$ cm$^{-3}$. 
The HI data are from Boulanger $\&$ Viallefond \cite{NGC6946HI} and Ryder \etal 
\cite{Ryder} for NGC6946 and NGC1313 respectively. The horizontal dashed lines  correspond to the
[CII] intensities found by Madden \etal \cite{Madden} for three components in NGC6946.
Also plotted in the same figure are  galaxies observed with the  KAO and presented
by Stacey \etal \cite{Stacey}. The KAO data have a FWHM beam equal to
55$\arcsec$, and Stacey looked at the nucleus of each galaxy, so those results are biased toward the
highest [CII] surface brightness. This explains why the squares in Fig. 9 are generally well above the
individual measurements inside NGC1313 and NGC6946. We note, however, that
the Stacey \etal \cite{Stacey} deprojected surface brightness value for the nuclear region
of NGC6946 agrees with the nuclear value from the present work to within the errors.

\subsubsection{[CII] emission and the molecular gas}
Stacey \etal \cite{Stacey} established observationally that [CII] intensity  
increases linearly with CO intensity for galaxies of similar type, and
that the I$_{[CII]}$/I$_{CO}$ ratio increases with the global star formation activity
of galaxies, as measured by dust temperature.
In particular,  galaxies with warm dust (T$_{dust}>$ 40 K) have mean 
I$_{[CII]}$/I$_{CO}$ ratios of $\sim$4200, while galaxies with relatively cold dust
(T$_{dust}<$ 40 K) dust have mean I$_{[CII]}$/I$_{CO}$ ratios of  
$\sim$1300. Moreover, the correlation of I$_{[CII]}$ with I$_{CO}$
followed by galaxies with warm dust agrees very well with that traced
by Galactic star forming regions and shows much less scatter than the 
correlation followed by cooler galaxies. Wolfire et al. \cite{Wolfire}
showed in model calculations that this correlation arises naturally in PDRs, where
[CII] emission comes from warm PDR surface layers while CO emission comes from cooler 
UV-shielded gas deeper into PDRs. Kaufman et al. \cite{Kaufman} extended these
theoretical calculations to the case of low metallicity galaxies.

The [CII]--CO correlation holds for regions inside NGC6946 and NGC1313.
In particular, the sequence traced by NGC6946 agrees with that of the ISO--KP
sample of normal galaxies. We note, however, that the mean I$_{[CII]}$/I$_{CO}$
ratio for the regions inside NGC6946 is lower ($\sim$ 550) than the mean value found by Stacey
for the normal spirals in their sample, while most of the
ISO--KP galaxies agree with the measurements of regions in NGC6946.
This difference can be explained by the fact that the flux in the Stacey \etal 
\cite{Stacey} sample comes principally from the 55$\arcsec$ central region of the 
galaxies whereas the ISO--KP sample data refer to the
integrated emission of the galaxies (Malhotra \etal \cite{Malhotra}).
The Stacey \etal \cite{Stacey} sample is biased towards more active regions, even when the galaxies 
are relatively cool, normal galaxies.
The integrated flux of the ISO--KP galaxy sample (Tables \ref{tab4} and
\ref{tab7}) and the regions inside the disk
of NGC6946 are likely to come from  a mixture of star forming and more quiescent regions.

Among the three regions observed and detected in NGC1313 in CO at 115 GHz, only one 
(region 69 in Tables \ref{tab2} and \ref{tab8})
has a I$_{[CII]}$/I$_{CO}$ ratio in agreement with the mean value found in
NGC6946 and for the ISO--KP galaxies.
The other two regions have much higher I$_{[CII]}$/I$_{CO}$ ratios, similar
to the ratio found in more active galaxies and galactic star forming regions. 
However, although 
the [CII] intensity of these two regions is high compared to other point in this galaxy, 
the [CII] intensities are low compared to most active galaxies and star forming regions.
The CO intensity is exceptionally low, leading to the high [CII]/CO ratios. Even more telling 
is that the dust temperatures, as measured by the 60$\mu$m/100$\mu$m ratio, are low. 
Therefore, the
high [CII]/CO ratios in NGC1313 likely have a completely different explanation than 
the fact that they arise in
active  star forming regions. We propose 
that in NGC1313, the high 
ratio is explained by [CII] emission arising from low extinction diffuse gas, where
 penetration by FUV photons prevents the formation of CO. The low metallicity of 
 NGC1313 may contribute to this effect, 
by lowering the dust extinction. In low metallicity 
environments, CO molecular clouds can be  smaller and surrounded by
 large  CII envelopes   (Lequeux \etal \cite{Lequeux}, Pak \etal \cite{Pak}, Kaufman \etal \cite{Kaufman}), 
resulting in large [CII]/CO ratios.
It is in fact this scenario which is used to explain the very high [CII]/CO
intensity ratio seen in 30 Dor (see Fig. 11 of Kaufman et al. 1999).

What then explains the lower I$_{[CII]}$/I$_{CO}$ ratio in the one beam out of three? 
One possibility is that there is a significant variation of metal content   
within NGC1313: region n.69 may have an higher metal content than regions 89 and
90. This explanation is unlikely because  no such  metallicity 
variation has been observed in previous studies (Moll\'a $\&$ Roy
\cite{Molla}). Another possible explanation is  
that it is not metallicity which drives the observed differences in the 
[CII]/CO ratios but the total hydrogen column densities of the clouds. Kaufman \etal
\cite{Kaufman} discussed this point in detail. A cloud in a low metallicity environment
may have a thin [CII] shell surrounding a large  CO core, resulting in
[CII]/CO ratios comparable to those observed in normal metallicity environments,
if A$_V$ through the cloud has the typical
value of normal metallicity clouds (A$_V$$\sim$5-10) in our Galaxy. In
other words, the observations are consistent with the clouds in the beam (reg. 
69) having 
higher columns
on average than the clouds in the beams of regions 89 and 90.  However,
the  total {\it beam-averaged} hydrogen nucleus column densities, 
N(HI)+2$\times$N(H$_2$), are all 
comparable in the three regions (Table \ref{tab9}). This then implies that 
there are fewer
clouds in the beam of region 69 than in the beams of regions 89 and 90,
so that the individual clouds columns in region 69 are higher than in
regions 89 and 90, even though the beam-average column of the ensemble
of clouds is similar in all three regions.

An other interesting conclusion
is that the physical conditions in NGC1313 seem to be much more inhomogeneous than
those in the disk of NGC6946. Regions 69 and 91 have comparable [CII]
emission but very different molecular content: region 91 has not been detected
at SEST. Regions 89 and 90 have   the
highest  [CII] emission and very little molecular gas. The 
[CII]/CO ratio decreases  along the S-shape region outlined by
star-forming regions in NGC1313, going
from the south to north-east. 
Wells \etal \cite{Wells} also reported gradients along this region,
with each of infrared-to-radio, infrared excess, and 60-to-100$\mu$m color decreasing.
This could reflect differences in the star
formation properties along the spiral arms of the galaxy, or as suggested
by Wells \etal \cite{Wells} a decreasing dust-to-gas ratio. A more detailed
analysis 
on this issue, which is beyond the aim of this paper,  requires further CO 
observations and a comparison with the
star-formation properties of the different regions inside NGC1313.

\subsubsection{The [CII] emission and the HI gas}

In general, we would expect a correlation between HI and [CII] if
both arises in PDRs.
However, as discussed in the introduction, there are basically two types of PDRs which 
contribute significantly to neutral emission from a galaxy:
low density diffuse gas illuminated by the general ISRF (G$_0\sim$1) and 
the dense ($n\gtrsim n_{crit}$) and warm (G$_0\gg$ 1) surfaces of UV illuminated Giant Molecular Clouds 
(GMCs). Thus, we would only expect a correlation between [CII] and HI if the bulk of the
emission from an observed region comes from the same type of PDR, or the same admixture of phases.

Figure 10 shows that in NGC6946 the [CII] intensity spans more than one order of magnitude regardless
of the HI intensity (or column density if the emission is optically thin). The absence of a 
correlation can arise if one or more of the following conditions are satisfied:

1) If most of the HI in the LWS beam is optically thick, we are 
underestimating the HI column density; the HI column may vary but produce 
relatively constant HI intensity. As the total column increased, the
[CII] intensity would increase, but the HI intensity would remain
at a value corresponding to the emission from optically thin gas
with a column given by HI optical depth of order unity.
The average HI column density, based on the assumption of optically thin 
emission, of regions in NGC6946 with [CII] intensities
higher than the intensity expected from standard HI clouds, is
$\sim$7$\times$10$^{20}$ cm$^{-2}$. The HI opacity is given by: 
\begin{equation}
\tau_{HI} = {\frac{N(HI)}{1.83\times10^{18} T \Delta v_5}},
\end{equation}
where, $\Delta v_5$ is the FWHM of the line at 21 cm from a single cloud in 
units of km s$^{-1}$, and N(HI) is in units of cm$^{-2}$ for a single cloud.
 The HI lines in NGC6946 are 
broader than $\sim$ 8 km s$^{-1}$ (Boulanger $\&$ Viallefond 
\cite{NGC6946HI}).  If the emission arises from a single cloud
which fills the beam, this value, the observed HI column, and an 
estimated temperature of about 50 K{\footnote{  Boulanger $\&$ Viallefond (
\cite{NGC6946HI})
found   temperatures of the HI clouds in  4 annuli centered on the galaxy
lower than 50 K.  But these values
represent the average on a much larger area than the LWS beam.  Note also that 
the typical temperatures at the surface of 
the PDRs found applying the model to our observations are much higher than 50
K.}}
results in a value of $\tau \sim 1$.  However, it is more likely that
the beam encompasses a number of clouds.  Individual clouds in
the Milky Way have smaller velocity dispersions, 
$\Delta v$ $\approx$ 2 km s$^{-1}$. 
It would take a minimum of 4 of such clouds to produce the observed
FWHM of the line of 8 km s$^{-1}$.  If each cloud filled the beam,
each cloud could only contribute to the ``observed column" 
about 1.75$\times 10^{20}$ cm$^{-2}$.   Again,
using the FWHM from a single cloud and the ``observed column" of a single
cloud, along with a temperature of 50 K, results in $\tau \sim 1$.
Therefore, if these presumably diffuse clouds have  temperatures close to
 50 K, it is consistent that there may be a very large total column in the
clouds to make the HI line very optically thick, and saturate the HI
emission at a level consistent with the assumed optically thin
 column of $7\times 10^{20}$ cm$^{-2}$.  However, if the temperature
is higher than about 50 K, the only way for this scenario to work
is to have the clouds only partially fill the beam.  This is highly
unlikely to explain the observations, since a large number of
observations at different points all gave the same HI intensity;
this would require the same beam filling factor in each beam.

Can optically thick HI clouds produce the observed wide range
in [CII] intensity?  The [CII] emission scales linearly with 
column density only if 
conditions are {\it effectively} optically thin; once [CII] emission is 
effectively thick, the intensity saturates at the blackbody value, increasing
only logarithmically with column as the line broadens. The optical depth of
the [CII] line is (Crawford \etal \cite{Crawford}):
\begin{eqnarray}
\tau_{[CII]}&=& 0.37\left[\left(1+\frac{n_{crit}}{n}\right) e^{(92/T)} -
1\right]\\\nonumber
& &\times \left( \frac{2e^{(-92/T)}}{1+2e^{(-92/T)}+n_{crit}/n}\right) 
x(C)_{-4}N(HI)_{21} \Delta v_5^{-1},
\end{eqnarray}
and 
\begin{equation}
\tau_{eff{\rm [CII]}}\approx \tau_{\rm [CII]}\bigg[\frac{1}{1+n_{crit}/n}\bigg],
\end{equation}
where N(HI)=10$^{21}$N(HI)$_{21}\,\rm cm^{-2}$ is the column density of hydrogen 
nuclei associated 
with the [CII] emission, and $x$(C)=10$^{-4}x$(C)$_{-4}$ is the abundance of 
carbon relative to H nuclei. 
For $n\gtrsim n_{cr}$ and $x(C)_{-4}=1.4$, $\tau_{CII}
=1$ for an HI column density $\sim 1.5\times 10^{21}\,\rm cm^{-2}$. 
Therefore, in these conditions, once the HI intensity
has saturated we would expect I[CII] in NGC6946 to increase
only by a factor 1.5$\times$10$^{21}$/7$\times$10$^{20}$ $\sim$ 2 before the 
[CII] line also 
becomes optically thick and saturates. 
On the other hand, if $n\lesssim 0.1n_{cr}\approx 3\times 10^2\,\rm cm^{-3}$, 
the 
[CII] line remains effectively thin up to hydrogen columns $\gtrsim 10^{22}\,\rm 
cm^{-2}$, and the
observed range of [CII] intensity with similar HI intensity could be reproduced. 
However, this would require at positions with the highest [CII] 
intensities columns $N(HI)\gtrsim  10^{22}\,\rm 
cm^{-2}$, 
through the disk 
of NGC6946, much greater than columns through the Milky Way. 
In addition, it does not directly explain why [CII] 
spatially 
correlates with CO and the FIR continuum but
not with HI (Figure 6).

2) Another possibility is that [CII] emission arises mostly from HII regions, while the HI emission comes from 
diffuse neutral gas. If this were the case, then as the [CII]-to-[NII] ratio decreases and approaches 
the value predicted for ionized gas ([CII]/[NII]$\sim 0.25$ for dense HII regions and $\sim 4.3$ for 
diffuse ionized gas) the [CII]-to-HI ratio should increase.
In Figure 11, we show that in NGC6946, as [CII]/[NII] decreases [CII]/HI increases, i.e. as more of 
the [CII] comes from ionized gas there appears to be more [CII] relative to HI. 
However, in the regions with exceptionally high [CII]/HI, this scenario would predict [CII]/[NII]$\lesssim 4.3$, whereas the 
observations show the ratio never dropping below $\sim 8$ in NGC6946.

3) If the bulk of [CII] in most of the observed regions in NGC6946 comes from
PDRs associated with dense gas but the bulk of HI comes from the diffuse neutral 
gas component, [CII] and N(HI) should not be correlated. This is consistent with the 
spatial anticorrelation of [CII] and HI shown in Figure 6. This suggests that HI comes from 
a diffuse component ($n\lesssim 10^3\,\rm cm^{-3}$) which produces insignificant [CII] emission,
whereas the [CII] emission arises from a dense neutral component that produces insignificant average
HI column in the beam. The dense component which produces [CII] should also produce the bulk of the 
CO and a significant amount of FIR emission.
As we have shown in the previous Section and in Fig. 9, this is exactly what we observe in NGC6946. 
Moreover, as we will discuss in \S 5.4, the  HI produced in dense (n$\gg$n$_{crit}$)
gas  associated with star forming regions, {\it  i.e} dense PDRs, is only a few 
$\%$  of the total HI, and its amount  does not vary significantly among the observed regions
in the disk of NGC6946.

This last scenario is the most likely: the increase of [CII] observed in NGC6946  (Fig. 9) 
is due to dense PDRs related to star forming regions, with a possible small contribution from 
the ionized gas associated with these star forming regions. This view is supported by the 
spatial cuts in Figure 6, which show that that HI and [CII] do not track across the plane of 
NGC6946. This is also corroborated by the PDR modelling results shown in Figure 7, indicating that 
the [CII], [OI] and FIR emission is fit by PDR gas with densities $\gtrsim 10^3\,\rm cm^{-3}$. 

The situation is not as clear in NGC1313. Here there is a gradual rise in [CII] 
intensity with increasing HI column up to $N(HI)\sim 10^{21}\,\rm cm^{-2}$,
consistent with HI and [CII] being correlated. Above $N(HI)\sim 10^{21}\,\rm cm^{-2}$, 
the [CII] intensity rises by a factor $\sim 7$, not an unreasonable 
amount if the [CII] emission remains effectively thin when the HI 21 cm line goes optically thick. 
Thus, a significant amount of the 
[CII] from NGC1313 could be coming from diffuse neutral gas. The fact that [CII] emission 
does not seem to go optically thick until higher columns of HI than in NGC6946 may also be due
to low metallicity in NGC1313 (see Kaufman \etal 1999). Unfortunately, there is insufficient
data to spatially correlate CO and [CII] in NGC1313 as we have done in NGC6946. If the same correlation
exists in NGC1313 as in NGC6946, it would be powerful evidence that [CII] arises from the 
dense PDR surfaces of the CO clouds. We note that Wolfire \etal \cite{Wolfire} showed that for reasonable 
interstellar cloud masses and size distributions, dense clouds 
will dominate the [CII] emission, a result reinforced by the correlation of 
[CII] and CO in many galaxies.  Lacking this evidence, however, we must consider whether
[CII] might arise mainly from diffuse gas in NGC1313.   

For the regions with low [CII] intensities    we point out 
that the HI  interferometric data fail to detect low surface brightness extended
emission.  Comparing the  NGC1313 HI total flux taken with the Australia Telescope Compact
Array (ATCA) and the Parkes 64m telescope, Ryder \etal \cite{Ryder} found that
6$\%$ of the total flux is lost with the interferometric observations.
If we assume that most of this missing flux comes from low surface brightness
extended emission in NGC1313, we conclude that the points in Fig. 9 with the
lower HI column density are probably underestimates. If we distribute this excess equally 
to those regions with $N(HI) < 10^{21}\,\rm cm^{-2}$,  the points in Fig. 9 between 10$^{20}$  
and 10$^{21}$ cm$^{-2}$ move up to between 5$\times$10$^{20}$ and 1.3$\times$10$^{21}$ cm$^{-2}$. This
makes the [CII]/HI points in NGC1313 move closer to the emission expected from 
low density (n$\sim$30 cm$^{-3}$) HI clouds. We conclude 
that in NGC1313 we successfully observe [CII] emission from two separate
ISM components. The first is associated with diffuse, low density 
(30 $<$ n $<$ 100 cm$^{-3}$) atomic gas probably illuminated by the general FUV
ISRF (standard atomic clouds), dominated by HI and contributing up to 
40$\%$ of the total [CII] emission, as seen from our comparison of [CII] and HI emission. 
The second component is associated with optically thick clouds near regions of star formation, as seen
from the three regions for which we have computed $G_0$ and $n$.

In conclusion, we find that in NGC6946, much of the [CII] emission comes from dense
PDRs on the surfaces of molecular clouds, the same clouds that are responsible for 
CO emission, while the HI emission appears to come from a more extended diffuse 
component. In NGC1313, some of the [CII] may come from low density atomic regions 
which also contribute to the observed HI emission; there is less CO emission 
associated with [CII] either because low metallicity leads to small CO cores surrounded
by large [CII] envelopes or because there are very few high extinction clouds relative
to diffuse clouds.

\subsection{Taking the models further: diffuse vs. dense HI gas} 

In our modelling to this point, we have attempted to fit the observed HI emission
with a single gas component: either HI gas associated with PDRs on molecular cloud
edges or from diffuse interstellar clouds. In addition, we have tried to explain the 
[CII], [OI] and FIR emission as arising from one component.
In this section, we estimate the contribution to HI column and the [CII] intensity
from each of the two neutral gas components. 
  
The origin of most of the atomic hydrogen in nearby galaxies is a controversial point.
The standard picture assumes that most of the HI in galaxies is a precursor to
star formation, and arises from a diffuse HI component. However, recent high spatial 
resolution observations of molecular, neutral and ionized gas in nearby galaxies point 
toward an evolutionary picture in 
which a sizeable fraction of the neutral hydrogen is a product of recent star formation, 
resulting from  photodissociation of H$_2$ in molecular clouds by the radiation emitted 
by young stars (Allen, Atherton $\&$ Tilanus \cite{Allen86},  Allen \etal
\cite{Allen97}, Smith \etal \cite{Smith}). This was also suggested by the
relative brightness of [CII] compared with  HI line emission from galaxies 
in the Stacey \etal \cite{Stacey} sample, which indicated that the [CII] and HI emission
arose from relatively dense gas.

We want to estimate how much of the HI in NGC6946 and NGC1313 is produced by recent 
photodissociation of H$_2$ at the PDR interfaces between HII regions and molecular clouds (the 
``dense'' component), and how much originates in diffuse gas heated by the general
radiation field.  The observed HI is: 
\begin{equation}
N(HI)_{obs}=N(HI)_{dense}+N(HI)_{diff},
\end{equation}
where the column densities are the averages of each component within the beam.
The HI diffuse component should in principle be separated into   emission
arising from Cold and Warm Neutral Medium (CNM, n$\sim$ 30 cm$^{-3}$, 
T$\sim$100 K, WNM, n$\sim$0.3 cm$^{-3}$, T$\sim$8000 K). The WNM does not 
contribute
significantly to the [CII] emission (Wolfire \etal \cite{Wolfire}) whereas it 
might significantly contribute 
the total HI diffuse emission. In what follows we neglect the WNM contribution 
to
the diffuse HI component, because we do not have sufficient constraints for
developing a three component model. Although a weakness of our model, this
approximation is
partially justified by observations inside and at the  solar circle of the Milky
Way galaxy that the 
CNM  HI(21 cm) emission is comparable to or dominates the WNM HI(21 cm)  
emission (Dickey and Lockman, \cite{Dickey}, Kulkarni and Heiles
\cite{Kulkarni}).
The WNM may dominate outer
regions of HI(21 cm) emission of normal galaxies, but we point out that the 
regions we modelled
are well inside  R$_{25}$ (see Fig. 1). Moreover, we  stress
that our  model represents an improvement, though imperfect,  over a one
component model.\\ 
An equation  similar to Eq. 5  holds for the observed [CII] flux coming from the neutral gas,
{\it i.e.} with the ionized gas contribution removed:
\begin{equation} 
I_{[CII]}^{PDR}=I_{[CII]}^{{PDR}_{dense}}+I_{[CII]}^{{PDR}_{diff}}. 
\end{equation}

In the optically thin regime, the column density of hydrogen nuclei associated
with a given [CII] intensity is (Crawford \etal \cite{Crawford})

\begin{equation}
N_{[CII]}(HI) =\frac{4.25\times 10^{20}}{x(C)}
\left[\frac{1+2e^{(-92/T)}+(n_{crit}/n)}{2e^{(-92/T)}}\right]~{I_{[CII]}^{PDR}},
\end{equation}

where $N_{[CII]}(HI)$ is in cm$^{-2}$, $I_{[CII]}^{PDR}$ in erg
s$^{-1}$ cm$^{-2}$~ sr$^{-1}$, and $x(C)$ is the C/H gas--phase   abundance ratio.

The column density of HI produced in dense gas (assuming $n=n_{dense}\gg n_{crit}$ for [CII]) is:
\begin{equation}
N(HI)_{dense} =\frac{4.25\times
10^{20}}{x(C)} \left[\frac{1+2e^{-92/T_{dense}}}{2e^{-92/T_{dense}}}\right]~I_{[CII]}^{{PDR}_{dense}}.
\end{equation}

and the column density of HI associated with diffuse neutral gas is:

\begin{equation}
N(HI)_{diff}=\frac{4.25\times
10^{20}}{x(C)}\left[\frac{1+2e^{-92/T_{diff}}+(n_{crit}/n_{diff})}{2e^{-92/T_{diff}}}\right]~I_{[CII]}^{{PDR}_{diff}}.
\end{equation}

Here, $T_{dense}$ and $T_{diff}$ are the temperatures of the dense and diffuse components, respectively.

In the limit where the density of the diffuse gas $n_{diff}$ is much less than the critical density for the
[CII] transition ($n_{crit}\simeq 3\times 10^3\,\rm cm^{-3}$), 
we can solve for the [CII] intensity arising from the dense gas only:

\begin{equation}
I_{[CII]}^{{PDR}_{dense}}=I_{[CII]}^{PDR}-\frac{2x(C)N(HI)_{obs}}{4.25\times 10^{20}e^{92/T_{diff}}}\frac{n}{n_{crit}}
\end{equation}

where the second term on the right hand side is the [CII] intensity from the 
diffuse gas (see Appendix B for detailed calculation). For 
$T_{diff}\sim 92~K$ and taking a diffuse component density of $30\rm\,cm^{-3}$, 
\begin{equation}
I_{[CII]}^{{PDR}_{dense}}=I_{[CII]}^{PDR}-1.73\times 10^{-6}x(C)_{-4}N(HI)_{21},
\end{equation}
where $N(HI)_{21}=N(HI)_{obs}/10^{21}\,\rm cm^{-2}$. 
Thus, if we know $n$, the density in the diffuse gas, we can separate the [CII] emission from the
dense and diffuse components. Finally, the column of HI arising from dense gas is:
\begin{equation}
N(HI)_{dense}=\frac{6.4\times 10^{24}}{x(C)_{-4}}\left[I_{[CII]}^{PDR}-1.73\times 10^{-6}x(C)_{-4}N(HI)_{21}\right].
\end{equation}

Tables \ref{tab10} and \ref{tab11} show the results of this calculation for NGC6946 and NGC1313
respectively in cases where the observed [CII] emission has been corrected for 
the contribution from ionized gas.
In NGC6946 most of the [CII] emission observed in neutral gas, 
arises in dense PDRs (Col. 2 of table 9). This component seems to contribute less
(a factor of 2) in the three modelled region of NGC1313 (Col. 2 of table 10), confirming that the diffuse
neutral gas   contribute more in the disk of
NGC1313 than in NGC6946 to the total emission. 
Moreover, this result supports the conclusion in $\S$ 4.2, and the picture used to interpret the 
behaviour between [CII] intensity and HI column density presented in $\S$ 5.3.2.

The fraction of HI column density, $N(HI)_{dense}/N(HI)_{obs}$, coming from dense neutral gas in each region
is listed in Col. 3 of Tables \ref{tab10} and \ref{tab11}. We find typically that only a few percent of the HI column is associated
with dense neutral gas near star forming regions.   This result is in   
good agreement with  the values published by  Stacey \etal  \cite{Stacey} 
  for normal galaxies. In conclusion, not much of the HI can be coming from the densest
gas in spiral arms, where recent star formation has occurred. However, our spatial resolution is insufficient to 
determine whether the diffuse HI, which dominated the 21 cm production, arises from recently dispersed GMCs or
from an older, more pervasive interarm component. Whether much of 
galaxy's HI column is produced by recent photodissociation of H$_2$ in spiral arms remains an open 
question.

\section{Summary and Conclusions}
In this paper we  presented new ISO-LWS observations   in the main FIR fine
structure lines of the two nearby spiral galaxies: NGC1313 and NGC6946. 
Both galaxies were fully mapped in  the [CII (158 $\mu$m)] line with a linear resolution of $\sim$ 1.5 kpc. 
Some regions in NGC1313 and   in NGC6946 were also  observed 
in  the  [OI(63 $\mu$m)]) and the [NII(122 $\mu$m)]   lines with comparable resolution.

In  PDRs most of the carbon is ionized and most of
the oxygen is neutral.  
In these regions,  the atomic gas is mainly  heated by photo--electrons ejected by grains
after absorption of a FUV (6 eV $<$ h$\nu$ $<$ 13.6 eV) photon
and cools primarily {\it via} the [CII] and [OI] lines. Therefore,
a comparison of the dust emission at MIR and FIR wavelengths to the emission in
these cooling lines gives important clues on the 
heating/cooling processes of the atomic medium.\\  
Following this line of thought, Malhotra \etal \cite{Malhotra} and Helou \etal
\cite{CAFE} studied   how  the   integrated [CII] emission of a sample of
star--forming galaxies (ISO--KP) relate to the IR dust emission as a function
of the overall galaxy activity   as traced by the 60/100 $\mu$m IRAS colors. 
They found that the [CII]/FIR and [CII]/MIR ratio behave very differently.
The [CII]/FIR ratio decreases with the IRAS colors whereas the [CII]/MIR stays constant.  This
difference  has profound consequences in the understanding of the role that different
dust grain populations play in the atomic gas heating/cooling process. In particular, 
these results confirm and strengthen the theoretical prediction that, at least in normal
star--forming   galaxies, PAHs are the most efficient grain population 
for photo--electron production (Bakes $\&$ Tielens
\cite{Bakes}).
 
The analysis in NGC1313 and NGC6946 presented in this paper had three goals:\\
1) investigating whether the above behaviour, found for the integrated
emission of normal galaxies,   holds as well in regions associated 
with different galaxy components like nucleus, spiral arms and diffuse disk.\\
2) deriving the average physical conditions (gas density $n$, FUV radiation
field $G_0$ and surface temperature 
$T_s$) of the emitting clouds  in the NGC1313 and NGC6946 regions observed with 
ISO-LWS,   by applying   the PDR model predictions 
by Kaufman \etal \cite{Kaufman}.  The resulting parameters were also compared with the
corresponding parameters found for   the integrated emission of the ISO--KP galaxies.\\
3) comparing the observed [CII] emission in NGC1313 and NGC6946 to 
other gas component tracers like HI (21 cm) 
and $^{12}$CO(1--0) in order to investigate from which component of the neutral gas 
(ionized medium, low   and high density PDRs)   the  observed  [CII]
emission  inside NGC1313 and NGC6946 arises, and if there are differences 
between regions associated with different galaxy components.  \\
We  summarise the main results as follows:

  We found no statistical difference between the [CII]/FIR and [CII]/MIR
  ratios as a function of the 60/100 colors inside NGC1313 and NGC6946 and   
  the integrated ratios of ISO--KP sample. However, the [CII]/MIR average ratios
  inside NGC1313 and the integrated ratios of some others irregulars (Hunter \etal \cite{Hunter}) are
  systematically higher than the corresponding ratios in NGC6946. We show that at least
  inside NGC1313, this result is due to a deficiency of the carriers responsible
  for the Aromatic Features seen in emission at MIR wavelengths
     with respect to
  the aromatic carriers   in NGC6946.   
  This deficiency can arise either from an originally
  lower carbon-based grain production due to  a metallicity lower than the solar
  value, and/or to an enhanced destruction of carriers responsible for AFE from the intense and hard 
  interstellar radiation field.\\
We did not find high ($>$ 0.6) IRAS 60/100 $\mu$m colors 
 even in those regions associated with HII regions or the nuclei,
probably due to the relatively large beam size ($\sim$1.5 kpc) diluting
the emission from the most active regions with contributions from cooler dust.

  The total [CII] emission is   $\sim$ 0.8$\%$ and $>$0.5$\%$ of the FIR emission
  in NGC6946 and NGC1313, respectively.  Our estimation of the [CII] emission from
  different galaxy components  leads to the result that the diffuse component 
  associated with the disk of the galaxies is less than that found in other work
  (Madden \etal \cite{Madden}).
  In particular, we find that this component accounts for $<$ 40 $\%$ in NGC6946 and
  $\sim$ 30 $\%$  in NGC1313.
  
  We applied the   PDR model predictions (Kaufman \etal
  \cite{Kaufman}) to those regions of NGC6946 and NGC1313 
  observed in [OI], [CII] and [NII] (12 and 3 respectively), after removal of the [CII] emission arising in
  ionized gas. In this way we were able to derive  the average parameters 
  ($n$, $G_0$ and T$_s$) of the  neutral atomic gas in the LWS beam. 
  Although  these values roughly  agree with the 
  parameters found for the integrated emission of the ISO--KP sample (Malhotra \etal
  {\cite{Malhotra}), we find a   scatter in the $G_0$/$n$ ratio distribution
  much greater   than for the ISO--KP galaxies. This variation is
  probably related to differences in the stellar populations and/or to differences
  in the mixture of active
  and not active components inside the LWS beam, along the galaxies.\\
    We do not find regions as extreme ($n>10^4$ cm$^{-3}$ and $G_0>10^4$) as some ISO--KP
  galaxies. Though  
  surprising, this agrees with the fact that we did not find regions with 
  high 60/100 ratios. We attribute this result to   the fact that the 
  typical region within the beam (1.5 kpc) is still too big, leading to
a mixture of emission from active and less active regions.
   
In NGC6946,  [CII] and  $^{12}$CO(1--0) are well correlated, though the average
I$_{[CII]}$/I$_{CO}$ ratio is lower (a factor of  $\sim$ 2) than that 
previously found by Stacey \etal \cite{Stacey} for normal galaxies. The same trend is followed by the
integrated ratios of the ISO--KP sample for which CO data are available.
Since higher ratios correspond to
higher star formation activity, this shows  that the observed [CII] emission
in NGC6946 arises in   regions less active than in the galaxies observed by
Stacey \etal \cite{Stacey}, whose observations are biased
towards the   galaxy nuclei. \\
In NGC1313 we have $^{12}$CO(1--0) data for only three regions.   
One of these follows the trend outlined by the NGC6946 and ISO--KP 
observations, though at the very low surface brightness end of the relation. 
The other two regions have a I$_{[CII]}$/I$_{CO}$ in agreement with that expected for
more active regions even though they do not have particularly high 60/100
$\mu$m. We interpret this result as due either to a change in metallicity
along the galaxy (unlikely) or to a diffuse/dense gas ratio in the LWS beam
higher than in NGC6946.   Moreover,  
regions with similar [CII] emission have very different CO content, suggesting that
the physical condition of the ISM inside NGC1313 are   
much less homogeneous than   in NGC6946.
   
In NGC6946 the [CII] intensity spans more than one order of magnitude, 
and is independent of the HI column density. We have shown that this
behaviour is mostly due to the fact that [CII] and HI
arise from different gas components: the former is associated with 
PDRs in the interface between star forming regions and parental molecular
clouds, while the latter arises principally from diffuse, low density gas.\\
In NGC1313 the situation is very similar to that found in NGC6946 for HI column
densities $\gtrsim$ 10$^{21}$ cm$^{-2}$. In this case, however, clouds
optically thick in HI can explain at least part of the independence between
[CII] intensity and HI column density.  For HI column densities lower than
10$^{21}$ cm$^{-2}$ there is a weak proportionality, which corresponds
to what we expected from 
diffuse low density (n$\sim$30 cm$^{-3}$) PDRs illuminated by the general FUV interstellar
radiation field. Therefore, we successfully detected [CII]
emission arising from  two
distinct gas components in NGC1313: one is associated
with classical PDRs related to star forming and the other associated with 
the diffuse low density gas associated with  HI.
   
In agreement with the above results, 
we find that very little (a few $\%$)
of the observed HI flux originates in dense ($n>n_{crit}$$\simeq$3$\times$10$^3$
cm$^{-3}$) gas in the modeled regions of NGC6946. The same result holds for two
of the three regions modeled in NGC1313 (which correspond to location on the
galaxy associated to bright
star-froming complexes and not to the diffuse disk). In the  third region,
the observed [CII] emission might  
entirely   arise in diffuse ionized gas.
For all the other modeled regions (in both galaxies) the latter contribution
can be significant, but not more than $\sim$50$\%$.\\
On the other hand, most of the [CII] emission observed in NGC6946 arise
in dense gas associated to star forming regions whereas this component is a
factor $\sim$2 less in the three modelled region of NGC1313. This confirms
that   the diffuse  atomic gas disk component contributes more in NGC1313 than in
NGC6946 to the total [CII] observed emission.

\newpage

\appendix
\section{The foreground emission estimation at 158 $\mu$m in the NGC1313
direction}
To evaluate the [CII] foreground contribution for NGC1313, we calculated the
 [CII]  flux expected for an   IRAS 100 $\mu$m flux equal to the average emission 
around NGC1313 in the following way. 
To calculate the expected [CII]
emission for a given infrared (IR) flux we have to know which ISM phase
produces the observed IR emission: cirrus or molecular clouds.
Using the FIRAS and  Leiden/Dwingeloo HI data, Boulanger \etal
\cite{cirrusHI}
 presented the galactic IR-HI
correlation in the northern hemisphere of the Milky--Way.
The slope of this correlation for W$_{HI}$ $<$ 250 K km s$^{-1}$
(N(HI)$<$4.5$\times$10$^{20}$ cm$^{-2}$)  is 0.53.
At higher column densities the correlation  becomes steeper because of the increasing
contribution of the molecular component.
We use the Colomb, Poppel $\&$ Heiles \cite{Colomb} HI observations of the southern sky
to check if the IR emission around NGC1313 follows the cirrus relation.
Fourteen HI measurements at  30$\arcmin$   resolution  were found in the 
2.5$^0$$\times$2.5$^0$ region centered on NGC1313. We calculate the IRAS emission at 100 $\mu$m
using a 30$\arcmin$ FWHM beam size in the same HI positions.
According to the DIRBE Explanatory Supplement:\\

\begin{equation}
FIRAS(100~ \mu m) =0.72 \times IRAS(100~  \mu m).
\end{equation}
We thus fitted a line to the HI--DIRBE relation after multiplying the IRAS flux
by 0.72.
The fitted slope is 0.4, as compared with 0.53 obtained from Boulanger \etal
\cite{cirrusHI}.
Moreover, all the HI pointings have HI column density less than 4.5 $\times$10$^{20}$
cm$^{-2}$.
We thus are confident that the bulk of the infrared emission around NGC1313 is
from cirrus.

There is not a measured IRAS 100 $\mu$m -- [CII]
surface brightness relation for cirrus in our Galaxy. To calculate it we
combined the DIRBE 100 $\mu$m -- N(HI) relation published by Boulanger \etal 
\cite{cirrusHI}:
\begin{equation}
I^{DIRBE}_{100~ \mu m}(MJy/sr) =0.53 \times N(HI) (10^{20} cm^{-2}),
\end{equation}
 and the  [CII] surface brightness -- N(HI) relation evaluated by Bennett \etal 
\cite{Bennett}   for cirrus:
\begin{equation}
I_{C^+} (10^{-6} erg s^{-1} cm^{-2} sr^{-1}) = (2.56 \times N(HI) (10^{20} cm^{-2}))/4 \pi .
\end{equation}

Combining these equations and taking into account also eq. A1 
we obtain the following  IRAS (100 $\mu$m) --[CII] relation for cirrus:
\begin{equation}
I_{C^+}(erg s^{-1} cm^{-2} sr^{-1})) = 0.268\times10^{-6} I_{100~ \mu m}(MJy/sr) .
\end{equation} 

The average  IRAS 100 $\mu$m flux around NGC1313 has been obtained on the  
HiRes IRAS image by calculating the flux in the LWS beam in regions located
 randomly
around the galaxy. The average emission at 100 $\mu$m thus obtained is equal 
to  is 2.76 MJy/sr, which implies a [CII]
surface brightness equal to  0.79$\times$10$^{-6}$ erg s$^{-1}$ cm$^{-2}$ sr$^{-1}$). This  
corresponds to 11.2$\times$10$^{-14}$ erg s$^{-1}$ cm$^{-2}$.
By contrast, the [CII] emission measured with LWS in the reference point 
is 7$\times$10$^{-14}$ erg s$^{-1}$ cm$^{-2}$.

\section{Detailed calculation for separating the dense and diffuse components in
PDRs}

In this section we perform the detailed calculation which leads to Eq. (10)
of Sec. 5.4.
From Eq. 9 we isolate $I_{[CII]}^{PDR_{diff}}$:

\begin{equation}
I_{[CII]}^{PDR_{diff}}= \frac{N(HI)_{diff}~x(C)}{4.25\times 10^{20}}~
\left[\frac{2e^{(-92/T_{diff})}}{1+2e^{(-92/T_{diff})}+(n_{crit}/n_{diff})}\right]~
\end{equation}

From Eq. 6:

\begin{equation}
I_{[CII]}^{PDR_{dense}}=I_{[CII]}^{PDR_{obs}}-I_{[CII]}^{PDR_{diff}}
\end{equation}

Substituting Eq. B1 in Eq. B2:

\begin{equation}
I_{[CII]}^{PDR_{dense}}=I_{[CII]}^{PDR_{obs}}-\frac{N(HI)_{diff}~x(C)}{4.25\times 10^{20}}~
\left[\frac{2e^{(-92/T_{diff})}}{1+2e^{(-92/T_{diff})}+(n_{crit}/n_{diff})}\right]~
\end{equation}

From Eq. 5:

\begin{equation}
N(HI)_{diff}=N(HI)_{obs}-N(HI)_{dense}
\end{equation}

Substituting in Eq. B3:

\begin{eqnarray} 
I_{[CII]}^{PDR_{dense}} = & I_{[CII]}^{PDR_{obs}}-  \frac{x(C)}{4.25\times 10^{20}}~
\left[\frac{2e^{(-92/T_{diff})}}{1+2e^{(-92/T_{diff})}+(n_{crit}/n_{diff})}\right]\nonumber
\nl
                        & \left[N(HI)_{obs}-N(HI)_{dense}\right]     
\end{eqnarray}

Developing and taking to the left hand all terms referring to the dense 
component one obtains:

\begin{eqnarray} 
I_{[CII]}^{PDR_{dense}}-\frac{x(C)}{4.25\times 10^{20}}~
\left[\frac{2e^{(-92/T_{diff})}}{1+2e^{(-92/T_{diff})}+(n_{crit}/n_{diff})}\right]~
N(HI)_{dense} =  \nonumber \nl
I_{[CII]}^{PDR_{obs}}-  \frac{x(C)}{4.25\times 10^{20}}~
\left[\frac{2e^{(-92/T_{diff})}}{1+2e^{(-92/T_{diff})}+(n_{crit}/n_{diff})}\right]~N(HI)_{obs}\nonumber \nl
~
\end{eqnarray}

Let's call for simplicity  the right hand side of Eq. B6 A.
Substituting the expression for  $N(HI)_{dense}$ of Eq. 8 and factoring
$I_{[CII]}^{PDR_{dense}}$:

\begin{equation}
I_{[CII]}^{PDR_{dense}}~\left[1-\left(\frac{1+2e^{(-92/T_{dense})}}{1+2e^{(-92/T_{diff})}+(n_{crit}/n_{diff})}\right)~
\left(\frac{1}{e^{\left(-92/T_{dense}+92/T_{diff}\right)}}\right)\right]= A
\end{equation}

Since $n_{diff}\gg n_{crit}$ Eq. B7 can be approximated:

\begin{equation}
I_{[CII]}^{PDR_{dense}}~\left[1-\frac{n_{diff}}{n_{crit}}~\left({1+2e^{(-92/T_{dense})}}\right)~
\left(\frac{1}{e^{\left(-92/T_{dense}+92/T_{diff}\right)}}\right)\right]\approx A
\end{equation}
 
where we have now written explicitly the term A. 
For   T$_{diff}$$\sim$ 92 K   and T$_{dense}$$\gg$ 92 K the term
in the square brackets is $\approx$1.  Therefore Eq. B8 becomes:

\begin{equation}
I_{[CII]}^{PDR_{dense}}\approx I_{[CII]}^{PDR_{obs}}-  \frac{x(C)}{4.25\times 10^{20}}~
\left[\frac{2e^{(-92/T_{diff})}}{1+2e^{(-92/T_{diff})}+(n_{crit}/n_{diff})}\right]~N(HI)_{obs}
\end{equation}
 
which for $n_{diff}\ll n_{crit}$ gives:

\begin{equation}
I_{[CII]}^{PDR_{dense}}\approx I_{[CII]}^{PDR_{obs}}-  \left[\frac{x(C)}{4.25\times 10^{20}}~
 \frac{n_{diff}}{n_{crit}}~2e^{(-92/T_{diff})}\right] ~N(HI)_{obs}
\end{equation}
 
which is Eq. 10 of Sec. 5.4.

\newpage
This work was supported by ISO data analysis funding from the US National
Aeronautics and Space Administration, and carried out at the Infrared
Processing and Analysis Center and the Jet Propulsion Laboratory of the
California Institute of Technology.  ISO is an ESA project with instruments
funded by ESA member states (especially the PI countries: France, Germany,
the Netherlands, and the United Kingdom), and with the participation of
ISAS and NASA.\\
The authors thank the anonymous referee for his/her very useful comments and inputs,
which improved the paper.\\
The authors  also thank M.D. Thornley, 
T. Helfer  and M. Regan  of the BIMA Survey on Normal Galaxies (SONG) team
 for  kindly providing   the unpublished $^{12}$CO(1--0) data of NGC6946;
F. Boulanger and F. Viallefond for   the HI 21 cm data of NGC6946 and 
S.D. Ryder for the HI 21 cm data of NGC1313.\\
A. C. wish also to thank J. Lequeux and M. Gerin for  very useful
discussions.\\
M.R wishes to acknowledge support fro FONDECYT(Chile) grant
No 1990881 and FONDECYT (Chile) International grant No 7990042

}

\newpage

\begin{deluxetable}{cccccccccccc}
\tiny
\tablewidth{7.5in}
\tablenum{1}
\tablecaption{
NGC6946 and NGC1313 properties. 
Col. 1 : NGC names;
Col. 2 and  3 : R.A. and Dec. J2000;
Col. 5, 6 and 7: Semi major and minor axes at the optical (R$_{25}$) extent  in
arcmin and  heliocentric velocity (km/s) (RC3, de Vaucouleurs \etal
\cite{RC3}) ; 
Col. 8 : Distances computed in the Local Group  reference frame and 
 H$_0$=75 Km/s/Mpc; 
Col. 9 : Logarithm of the total FIR (from IRAS) over blue ratio (Table 1 of  Dale \etal
\cite{Dale00}); 
Col. 10 :  f$_{\nu}$(60~\micron)/f$_{\nu}$(100~\micron) (IRAS color ratio 
IRAS values computed on HiRes IRAS images, Table 4 of Dale \etal \cite{Dale00}); 
Col. 11 : f$_{\nu}$(60~\micron) (Jy) flux density.
}
 \tablehead{
\colhead{Name}&\colhead{R.A.~~~~~~Dec.}&\colhead{Morph.}&\colhead{a~~~~~~~~b}&\colhead{$V_{hel}$}&\colhead{Dist.}& 
\colhead{log(FIR/B)}&\colhead{F$_{60}$/F$_{100}$}& \colhead{F$_{60}$}&\nl
\colhead{}&\colhead{~J2000}&\colhead{}&\colhead{$\arcmin$}&\colhead{km/s}&\colhead{Mpc}&
\colhead{}&\colhead{}&\colhead{Jy}
}
\startdata
NGC1313  & 03:18:22.24~ -66:28:41.8  &  Sd &9.1~~~6.9 & 475 & 3.6 & -0.43 &0.47 &45.7 &\nl
NGC6946  & 20:34:51.22~ 60:09:17.5  & Scd & 11.5~~~9.8&48&4.5&-0.34 & 0.46 &167.7& \nl
\enddata
\label{tab1}
\normalsize
\end{deluxetable}

\begin{deluxetable}{lcccccc}
\small
\tablewidth{7.5in}
\tablenum{2}
\tablecaption{
LWS observations for NGC1313. Number of regions correspond to the numbers
in Fig. 1. Region 50 corresponds to the
galaxy center (Table 1). Errors (shown in parenthesis) derive from the line and baseline fitting and do not
include the 30$\%$ calibration uncertainties. To obtained the intrinsic emission
in the [CII] line first substract the foreground value (11$\times$10$^{-14}$
erg s$^{-1}$ cm$^{-2}$)  then multiply by the extended source correction factor ($\sim$0.6).
The LWS beam size is: $\pi$ (75\arcsec/2)$^2$/ln2=1.5$\times$10$^{-7}$ sterad.
}
\def\x{$\times$}
\tablehead{
\colhead{No.}&\colhead{R.A.~~~~~~Dec.}&\colhead{[CII] flux}&\colhead{No.}&\colhead{R.A.~~~~~~Dec.}&\colhead{[CII] flux}&\nl
\colhead{}&\colhead{~~J2000}&\colhead{158~\micron}&\colhead{}&\colhead{~~J2000}&\colhead{158~\micron}&\nl 
\colhead{}&\colhead{h:m:s.s~~~~ \degree:\arcmin:\arcsec }&\colhead{
10$^{-14}$ erg s$^{-1}$ cm$^{-2}$}&\colhead{}&\colhead{h:m:s.s~~~~\degree:\arcmin:\arcsec }&\colhead{ 10$^{-14}$ erg s$^{-1}$ cm$^{-2}$}&\nl
}
\startdata
 1 ~~~ & 3:16:59.0~ -66:29:57.8 &  $<$38      &    45~~~ & 3:18:16.3~ -66:32:11.0 &  50~(5)	& \nl
 2 ~~~ & 3:17:09.3~ -66:29:15.0 &  $<$43      &    46~~~ & 3:18:17.9~ -66:25:55.9 &  31~(7)	& \nl
 3 ~~~ & 3:17:06.3~ -66:30:59.0 &  30~(5)     &    47~~~ & 3:18:19.2~ -66:30:27.0 &  124~(15)	& \nl
 4 ~~~ & 3:17:13.4~ -66:32:00.2 &  38~(6)     &    48~~~ & 3:18:20.6~ -66:34:56.6 &  $<$30	& \nl
 5 ~~~ & 3:17:16.6~ -66:30:15.8 &  44~(8)     &    49~~~ & 3:18:20.7~ -66:24:11.9 &  29~(5)	& \nl
 6 ~~~ & 3:17:19.5~ -66:28:31.8 &  45~(5)     &    50~~~ & 3:18:22.2~ -66:28:41.9 &  222~(11)	& \nl
 7 ~~~ & 3:17:20.8~ -66:33:01.8 &  43~(5)     &    51~~~ & 3:18:23.5~ -66:33:12.6 &  56~(6)	& \nl
 8 ~~~ & 3:17:23.7~ -66:31:17.7 &  40~(5)     &    52~~~ & 3:18:25.1~ -66:26:57.8 &  48~(5)	& \nl
 9 ~~~ & 3:17:26.8~ -66:29:32.6 &  53~(9)     &    53~~~ & 3:18:26.6~ -66:31:27.5 &  58~(6)	& \nl
 10~~~ & 3:17:27.9~ -66:34:03.0 &  27~(4)     &    54~~~ & 3:18:28.1~ -66:25:12.7 &  37~(6)	& \nl
 11~~~ & 3:17:29.8~ -66:27:48.6 &  34~(4)     &    55~~~ & 3:18:29.4~ -66:29:43.4 &  101~(5)	& \nl
 12~~~ & 3:17:31.1~ -66:32:18.6 &  30~(7)     &    56~~~ & 3:18:30.9~ -66:34:13.1 &  51~(6)	& \nl
 13~~~ & 3:17:34.0~ -66:30:34.5 &  37~(11)    &    57~~~ & 3:18:31.0~ -66:23:28.7 &  32~(8)	& \nl
 14~~~ & 3:17:35.3~ -66:35:04.2 &  24~(5)     &    58~~~ & 3:18:32.4~ -66:27:58.3 &  108~(9)	& \nl
 15~~~ & 3:17:37.1~ -66:28:49.4 &  43~(6)     &    59~~~ & 3:18:33.7~ -66:32:29.0 &  31~(6)	& \nl
 16~~~ & 3:17:38.2~ -66:33:20.1 &  27~(8)     &    60~~~ & 3:18:35.2~ -66:26:14.3 &  28~(4)	& \nl
 17~~~ & 3:17:39.9~ -66:27:05.4 &  50~(11)    &    61~~~ & 3:18:36.8~ -66:30:43.9 &  68~(10)	& \nl
 18~~~ & 3:17:41.3~ -66:31:35.0 &  67~(9)     &    62~~~ & 3:18:38.3~ -66:24:29.1 &  45~(6)	& \nl
 19~~~ & 3:17:42.4~ -66:36:05.4 &  23~(4)     &    63~~~ & 3:18:39.6~ -66:28:59.9 &  156~(1)	& \nl
 20~~~ & 3:17:44.2~ -66:29:51.3 &  63~(8)     &    64~~~ & 3:18:41.1~ -66:22:45.1 &  $<$38	& \nl
 21~~~ & 3:17:45.6~ -66:34:21.0 &  26~(5)     &    65~~~ & 3:18:41.1~ -66:33:29.5 &  36~(6)	& \nl
 22~~~ & 3:17:47.3~ -66:28:06.2 &  36~(7)     &    66~~~ & 3:18:42.6~ -66:27:14.7 &  43~(6)	& \nl
 23~~~ & 3:17:48.4~ -66:32:37.0 &  56~(9)     &    67~~~ & 3:18:44.0~ -66:31:45.4 &  42~(4.)	& \nl
 24~~~ & 3:17:49.8~ -66:37:06.9 &  27~(6)     &    68~~~ & 3:18:45.4~ -66:25:30.7 &  26~(5)	& \nl
 25~~~ & 3:17:50.2~ -66:26:22.1 &  48~(8)     &    69~~~ & 3:18:47.0~ -66:30:00.3 &  61~(5)	& \nl
 26~~~ & 3:17:51.6~ -66:30:51.8 &  62~(7)     &    70~~~ & 3:18:48.4~ -66:23:45.2 &  34~(9)	& \nl
 27~~~ & 3:17:52.7~ -66:35:22.9 &  38~(4)     &    71~~~ & 3:18:49.8~ -66:28:16.3 &  47~(5)	& \nl
 28~~~ & 3:17:54.4~ -66:29:07.8 &  80~(8)     &    72~~~ & 3:18:51.3~ -66:32:45.9 &  53~(7)	& \nl
 29~~~ & 3:17:55.8~ -66:33:37.8 &  36~(6)     &    73~~~ & 3:18:52.8~ -66:26:31.2 &  38~(6)	& \nl
 30~~~ & 3:17:57.5~ -66:27:22.7 &  48~(7)     &    74~~~ & 3:18:54.2~ -66:31:01.9 &  42~(7)	& \nl
 31~~~ & 3:17:58.7~ -66:31:53.8 &  57~(10)    &    75~~~ & 3:18:55.6~ -66:24:47.1 &  26~(6)	& \nl
 32~~~ & 3:18:00.1~ -66:36:23.4 &  $<$50      &    76~~~ & 3:18:57.2~ -66:29:16.8 &  45~(10)	& \nl
 33~~~ & 3:18:00.4~ -66:25:39.0 &  34~(6)     &    77~~~ & 3:19:00.0~ -66:27:32.7 &  49~(5)	& \nl
 34~~~ & 3:18:01.8~ -66:30:08.6 &  226~(12)   &    78~~~ & 3:19:01.6~ -66:32:02.4 &  44~(9)	& \nl
 35~~~ & 3:18:03.0~ -66:34:39.3 &  36~(11)    &    79~~~ & 3:19:02.9~ -66:25:47.2 &  42~(4)	& \nl
 36~~~ & 3:18:04.6~ -66:28:24.6 &  68~(9)     &    80~~~ & 3:19:04.3~ -66:30:18.3 &  45~(7)     & \nl
 37~~~ & 3:18:06.1~ -66:32:54.2 &  79~(9)     &    81~~~ & 3:19:07.4~ -66:28:33.2 &  41~(3)     & \nl
 38~~~ & 3:18:07.7~ -66:26:39.5 &  40~(7)     &    82~~~ & 3:19:10.1~ -66:26:49.2 &  37~(6)     & \nl
 39~~~ & 3:18:08.9~ -66:31:10.2 &  118~(8)    &    83~~~ & 3:19:11.8~ -66:31:18.8 &  30~(7)     & \nl
 40~~~ & 3:18:10.4~ -66:35:40.2 &  23~(3)     &    84~~~ & 3:19:14.6~ -66:29:34.8 &  31~(2)     & \nl
 41~~~ & 3:18:10.6~ -66:24:55.4 &  40~(8)     &    85~~~ & 3:19:17.5~ -66:27:49.3 &  40~(7)     & \nl
 42~~~ & 3:18:12.0~ -66:29:25.1 &  198~(15)   &    86~~~ & 3:19:22.0~ -66:30:35.3 &  25~(5)     & \nl
 43~~~ & 3:18:13.2~ -66:33:56.1 &  43~(4)     &    87~~~ & 3:19:24.7~ -66:28:51.2 &  40~(6)     & \nl
 44~~~ & 3:18:14.9~ -66:27:41.0 &  53~(7)     &    88~~~ & 3:19:32.0~ -66:29:51.3 &  36~(10)    & \nl

\enddata
\label{tab2}
\normalsize
\end{deluxetable}

\begin{deluxetable}{lccccccccccccc}
\small
\tablewidth{7.0in}
\tablenum{3}
\tablecaption{LWS line flux measurements  of three regions in NGC1313 observed in more than one FIR fine 
structure line. Region 89 is close to the galaxy's center.}
\def\x{$\times$}
\tablehead{
\colhead{No.}&\colhead{R.A.~~~~~~Dec.}&\colhead{[CII] flux}&\colhead{[OI] flux}&
\colhead{[NII] flux}&\colhead{[OIII] flux}&\nl
\colhead{}&\colhead{~~J2000}&\colhead{158~\micron}&\colhead{63~\micron}& 
\colhead{122~\micron}&\colhead{88~\micron}&\nl 
\colhead{}&\colhead{h:m:s.s~~~~~ \degree:\arcmin:\arcsec}&\colhead{ 10$^{-14}$
erg s$^{-1}$ cm$^{-2}$}&\colhead{ 10$^{-14}$ erg s$^{-1}$ cm$^{-2}$}&
\colhead{ 10$^{-14}$ erg s$^{-1}$ cm$^{-2}$}&\colhead{ 10$^{-14}$ erg s$^{-1}$ cm$^{-2}$}&\nl
}
\startdata
 89~~~  &3:18:26.9~ -66:28:36.5 &  237~(8)   &  113~(16)  & $<$22  &\nodata	 &  \nl
 90~~~  &3:18:09.0~ -66:29:59.6 &  206~(10)  &  133~(18)  & $<$15  & 164~(13)	   &  \nl
 91~~~  &3:18:04.2~ -66:32:29.4 &  79~(9)    &  44~(9)    & $<$20  &\nodata	  &  \nl
\enddata
\normalsize
\label{tab3}
\end{deluxetable}

\begin{deluxetable}{lccccccc}
\small
\tablewidth{7.5in}
\tablenum{4}
\tablecaption{ LWS observations for NGC6946. Numbers corresponds to the numbers
in Fig. 1. Region 31 corresponds to the galaxy
center. The foreground value is 31$\times$10$^{-14}$ erg s$^{-1}$ cm$^{-2}$.
}
\def\x{$\times$}
\tablehead{
\colhead{No.}&\colhead{R.A.~~~~~~Dec.}&\colhead{[CII]}&\colhead{[OI]}&\colhead{[OI] }&
\colhead{[NII]}&\colhead{[OIII]}&\nl
\colhead{}&\colhead{~~J2000}&\colhead{158~\micron}&\colhead{63~\micron}&\colhead{145~\micron}& 
\colhead{122~\micron}&\colhead{88~\micron}&\nl 
\colhead{}&\colhead{h:m:s.s~~~~~ \degree:\arcmin:\arcsec}&\colhead{10$^{-14}$
 erg s$^{-1}$ cm$^{-2}$}&\colhead{10$^{-14}$ erg s$^{-1}$ cm$^{-2}$}&\colhead{ 10$^{-14}$ erg s$^{-1}$ cm$^{-2}$}&
\colhead{10$^{-14}$ erg s$^{-1}$ cm$^{-2}$}&\colhead{10$^{-14}$ erg s$^{-1}$ cm$^{-2}$}&\nl
}
\startdata
1~~~~&  20:35:34.9~  60:02:57.8   &	$<$35	 &   \nodata	  &    \nodata        &    \nodata	&   \nodata	 & \nl
2~~~~&  20:34:00.3~  60:03:53.3   &	$<$54	 &   \nodata	  &    \nodata        &    \nodata	&   \nodata	 & \nl
3~~~~&  20:35:26.1~  60:04:13.1   &	$<$42	 &   \nodata	  &    \nodata        &    \nodata	&   \nodata	 & \nl
4~~~~&  20:34:48.3~  60:04:35.7   &	52~(4)   &   \nodata	  &    \nodata        &    \nodata	&   \nodata	 & \nl
5~~~~&  20:34:38.7~  60:04:41.5   &	117~(11) &   \nodata	  &    \nodata        &    \nodata	&   \nodata	 & \nl
6~~~~&  20:34:10.5~  60:04:57.7   &	$<$65	 &   \nodata	  &    \nodata        &    \nodata	&   \nodata	 & \nl
7~~~~&  20:35:17.3~  60:05:27.9   &	41~(7)   &   \nodata	  &    \nodata        &    \nodata	&   \nodata	 & \nl
8~~~~&  20:34:58.4~  60:05:40.2   &	69~(9)   &   \nodata	  &    \nodata        &    \nodata	&   \nodata	 & \nl
9~~~~&  20:34:48.9~  60:05:46.3   &	145~(15) &   \nodata	  &    \nodata        &    \nodata	&   \nodata	 & \nl
10~~~&  20:34:39.5~  60:05:51.7   &	108~(5)  &   57~(8)	  &    \nodata        &    \nodata	&   \nodata	 & \nl
11~~~&  20:34:30.0~  60:05:57.8   &	60~(11)  &   \nodata	  &    \nodata        &    \nodata	&   \nodata	 & \nl
12~~~&  20:34:20.7~  60:06:02.5   &	 $<$49   &   \nodata	  &    \nodata        &    \nodata	&   \nodata	 & \nl
13~~~&  20:35:08.7~  60:06:45.0   &	 139~(7) &   \nodata	  &    \nodata        &    \nodata	&   \nodata	 & \nl
14~~~&  20:34:59.1~  60:06:51.1   &	 153~(13)&   \nodata	  &    \nodata        &    \nodata	&   \nodata	 & \nl
15~~~&  20:34:49.6~  60:06:56.5   &	 304~(8) &   141~(10)	  &    \nodata        &    18~(3)	&   \nodata	 & \nl
16~~~&  20:34:40.1~  60:07:01.9   &	 172~(9) &   \nodata	  &    \nodata        &    \nodata	&   \nodata	 & \nl
17~~~&  20:34:30.8~  60:07:08.0   &	 111~(42)&   39~(8)	  &    \nodata        &    \nodata	&   \nodata	 & \nl
18~~~&  20:34:21.4~  60:07:14.2   &	 81~(11) &   \nodata	  &    \nodata        &    \nodata	&   \nodata	 & \nl
29~~~&  20:35:18.8~  60:07:49.8   &	 89~(7)  &   \nodata	  &    \nodata        &    \nodata	&   \nodata	 & \nl
20~~~&  20:35:09.3~  60:07:55.9   &	 211~(9) &   \nodata	  &    \nodata        &    \nodata	&   \nodata	 & \nl
21~~~&  20:34:59.9~  60:08:02.0   &	 394~(12)&   158~(14)	  &    \nodata        &    $<$38	&   66~(9)	 & \nl
22~~~&  20:34:50.3~  60:08:06.7   &	 492~(31)&   \nodata	  &    \nodata        &    \nodata	&   \nodata	 & \nl
23~~~&  20:34:41.0~  60:08:12.8   &	 357~(9) &   205~(18)	  &    \nodata        &    37~(10)	&   49~(10)	 & \nl
24~~~&  20:34:31.6~  60:08:18.9   &	 284~(13)&   \nodata	  &    \nodata        &    \nodata	&   \nodata	 & \nl
25~~~&  20:34:22.1~  60:08:24.0   &	 122~(5) &   45~(9)	  &    \nodata        &    \nodata	&   \nodata	 & \nl
26~~~&  20:34:12.7~  60:08:29.8   &	 42~(7)  &   \nodata	  &    \nodata        &    \nodata	&   \nodata	 & \nl
27~~~&  20:35:29.1~  60:08:54.2   &	 64~(4)  &   \nodata	  &    \nodata        &    \nodata	&   \nodata	 & \nl
28~~~&  20:35:19.6~  60:09:00.4   &	 224~(11)&   \nodata	  &    \nodata        &    \nodata	&   \nodata	 & \nl
29~~~&  20:35:10.1~  60:09:06.1   &	 500~(16)&   211~(26)	  &    \nodata        &    46~(7)	&   118~(9)	 & \nl
30~~~&  20:35:00.6~  60:09:11.1   &	 489~(14)&   \nodata	  &    \nodata        &    \nodata	&   \nodata	 & \nl
31~~~&  20:34:51.2~  60:09:17.6   &	1045~(21)&   606~(80)	  &    39~(7)	      &    107~(23)	&   175~(50)	 & \nl
32~~~&  20:34:41.8~  60:09:23.8   &	 354~(16)&   \nodata	  &    \nodata        &    \nodata	&   \nodata	 & \nl
33~~~&  20:34:32.3~  60:09:28.8   &	 222~(7) &   117~(11)	  &    \nodata        &    18~(3)	&   \nodata	 & \nl
34~~~&  20:34:22.8~  60:09:34.2   &	 164~(12)&   \nodata	  &    \nodata        &    \nodata	&   \nodata	 & \nl
35~~~&  20:35:29.7~  60:10:05.1   &	 144~(8) &   \nodata	  &    \nodata        &    \nodata	&   \nodata	 & \nl
36~~~&  20:35:20.3~  60:10:11.3   &	 311~(15)&   123~(17)	  &    \nodata        &    17~(2)	&   \nodata	 & \nl
37~~~&  20:35:10.8~  60:10:16.0   &	 401~(15)&   \nodata	  &    \nodata        &    \nodata	&   \nodata	 & \nl
38~~~&  20:35:01.4~  60:10:22.4   &	 447~(18)&   191~(15)	  &    \nodata        &    58~(11)	&   97~(15)	 & \nl
39~~~&  20:34:52.0~  60:10:28.5   &	 468~(14)&   \nodata	  &    \nodata        &    \nodata	&   \nodata	 & \nl
40~~~&  20:34:42.5~  60:10:33.6   &	 264~(8) &   136~(9)	  &    \nodata        &    17~(4)	&   43~(10)	 & \nl
41~~~&  20:34:33.0~  60:10:39.0   &	 297~(19)&   \nodata	  &    \nodata        &    \nodata	&    \nodata	 & \nl
42~~~&  20:35:21.0~  60:11:20.4   &	 131~(6) &   \nodata	  &    \nodata        &    \nodata	&    \nodata	 & \nl
43~~~&  20:35:11.6~  60:11:27.2   &	 304~(11)&   157~(17)	  &    \nodata        &    19~(6)	&   123~(15)	 & \nl
44~~~&  20:35:02.2~  60:11:33.0   &	 284~(10)&   \nodata	  &    \nodata        &    \nodata	&  \nodata	 & \nl
45~~~&  20:34:52.7~  60:11:38.4   &	 191~(7) &   56~(10)	  &    \nodata        &    12~(2)	&  \nodata	 & \nl
46~~~&  20:34:43.2~  60:11:43.8   &	 152~(13)&   \nodata	  &    \nodata        &    \nodata	&  \nodata	 & \nl
47~~~&  20:34:33.7~  60:11:49.9   &	 259~(11)&   110~(11)	  &    \nodata        &    11~(3)	&  \nodata	 & \nl
48~~~&  20:35:21.7~  60:12:32.4   &	 61~($<$)&   \nodata	  &    \nodata        &    \nodata	&  \nodata	 & \nl
49~~~&  20:35:12.4~  60:12:37.8   &	 61~(9)  &   \nodata	  &    \nodata        &    \nodata	&  \nodata	 & \nl
50~~~&  20:35:02.8~  60:12:43.2   &	 101~(7) &   $<$44	  &    \nodata        &    \nodata	&  \nodata	 & \nl
51~~~&  20:34:53.5~  60:12:48.6   &	 165~(10)&   \nodata	  &    \nodata        &    \nodata	&  \nodata	 & \nl
52~~~&  20:34:25.0~  60:13:05.9   &	 68~(4)  &   \nodata	  &    \nodata        &    \nodata	&  \nodata	 & \nl
53~~~&  20:35:31.9~  60:13:36.8   &	 35~($<$)&   \nodata	  &    \nodata        &    \nodata	&  \nodata	 & \nl
54~~~&  20:35:03.7~  60:13:53.0   &	 25~(6)  &   \nodata	  &    \nodata        &    \nodata	&  \nodata	 & \nl
55~~~&  20:34:16.4~  60:14:22.2   &	 43~(3)  &   \nodata	  &    \nodata        &    \nodata	&  \nodata	 & \nl
56~~~&  20:35:42.1~  60:14:41.6   &	 47~($<$)&   \nodata	  &    \nodata        &    \nodata	&  \nodata	 & \nl
\enddata				         
\normalsize
\label{tab4}
\end{deluxetable}

\begin{deluxetable}{ccccccccccccccccc}
\small
\tablewidth{7.0in}
\tablenum{5}
\tablecaption{PDR model solutions for regions in NGC6946 observed in the
[CII(158 $\mu$m],[OI(63 $\mu$m] and NII(122 $\mu$m] lines. 
Col. 1: Type of ionized medium (IM) contributing to the [CII] observed flux. 
Col. 2, 3 and 4: n, $G_0$/n  and T$_s$ model solutions for the   [CII] arising in PDR
only. 
Col. 5 Area filling factor for regions in PDR emitting the observed [CII]
contained in the LWS beam.  
Col. 6  : Percentage of [CII] emission coming from ionized gas.
}
\tablehead{
\colhead{~~~~~~~~~~~~~~~~~~~~~~~~~~~~~~~~~~~~~~}&\colhead{Type}&\colhead{n}&\colhead{$\frac{G_0}{n}$}& \colhead{T$_s$} &\colhead{$\Phi_A$} &
\colhead{IM} &\nl
\colhead{~~~~~~~~~~~~~~~~~~~~~~~~~~~~~~~~~~~~~~}&\colhead{}&\colhead{cm$^{-3}$}&\colhead{}&\colhead{K}&\colhead{}&\colhead{$\%$}&\nl
\colhead{~~~~~~~~~~~~~~~~~~~~~~~~~~~~~~~~~~~~~~}&\colhead{(1)}&\colhead{(2)}&\colhead{(3)}& \colhead{(4)} &\colhead{(5)} &\colhead{(6)} &\nl 
}
\startdata
      &          &        & NGC6946    &         &           &            & \nl
\hline
      &          &        & \bf{n. 15} &         &           &            & \nl
~~~~~~~~~~~~~~~~~~~~~~~~~~~~~~~~~~~~~~ &   DIM   &   1.8e3   &     0.2    &   229   & 0.02      &  28$\%$     & \nl
~~~~~~~~~~~~~~~~~~~~~~~~~~~~~~~~~~~~~~ &   HII   &   3.1e3   &     0.2    &   191   & 0.04      &  1.6$\%$    & \nl
      &          &        & \bf{n. 21} &         &           &            & \nl
~~~~~~~~~~~~~~~~~~~~~~~~~~~~~~~~~~~~~~ &   DIM   &   1.8e3   &     0.5    &   307   & 0.02      & $< $35$\%$  & \nl
~~~~~~~~~~~~~~~~~~~~~~~~~~~~~~~~~~~~~~ &   HII   &   1.0e3   &     0.3    &   234   & 0.04      & $< $2.0$\%$ & \nl
      &         &         & \bf{n. 23} &         &           &            & \nl
~~~~~~~~~~~~~~~~~~~~~~~~~~~~~~~~~~~~~~ &  DIM    &   3.1e3   &     0.3    &   300   & 0.01      & 48$\%$      & \nl
~~~~~~~~~~~~~~~~~~~~~~~~~~~~~~~~~~~~~~ &  HII    &   1.0e3   &     0.3    &   234   & 0.04      & 2.8$\%$     & \nl
      &         &         & \bf{n. 29} &         &           &            & \nl
~~~~~~~~~~~~~~~~~~~~~~~~~~~~~~~~~~~~~~ &  DIM    &   1.0e3   &     1.0    &   320   & 0.02      & 42$\%$      & \nl
~~~~~~~~~~~~~~~~~~~~~~~~~~~~~~~~~~~~~~ &  HII    &   3.1e2   &     1.0    &   270   & 0.09      & 2.4$\%$     & \nl
      &         &         & \bf{n. 31} &         &           &            & \nl
~~~~~~~~~~~~~~~~~~~~~~~~~~~~~~~~~~~~~~ &  DIM    &   1.8e3   &     0.5    &   307   & 0.05      & 45$\%$      & \nl
~~~~~~~~~~~~~~~~~~~~~~~~~~~~~~~~~~~~~~ &  HII    &   1.0e3   &     0.3    &   234   & 0.12      & 2.6$\%$     & \nl
      &         &         & \bf{n. 33} &         &           &            & \nl
~~~~~~~~~~~~~~~~~~~~~~~~~~~~~~~~~~~~~~ &  DIM    &   3.1e4   & $<$ 0.1    &  $<$100 & $>$1      & 41$\%$      & \nl
~~~~~~~~~~~~~~~~~~~~~~~~~~~~~~~~~~~~~~ &  HII    &   1.0e3   &     0.3    &   234   & 0.02      & 2.4$\%$     & \nl
      &         &         & \bf{n. 38} &         &           &            & \nl
~~~~~~~~~~~~~~~~~~~~~~~~~~~~~~~~~~~~~~ &  DIM    &   3.1e3   &     1.0    &   347   & 0.01      & 60$\%$      & \nl
~~~~~~~~~~~~~~~~~~~~~~~~~~~~~~~~~~~~~~ &  HII    &   5.6e2   &     0.5    &   245   & 0.06      & 3.5$\%$     & \nl
      &         &         & \bf{n. 40} &         &           &            & \nl
~~~~~~~~~~~~~~~~~~~~~~~~~~~~~~~~~~~~~~ &  DIM    &   1.8e3   &     0.1    &   221   & 0.02      & 32$\%$      & \nl
~~~~~~~~~~~~~~~~~~~~~~~~~~~~~~~~~~~~~~ &  HII    &   1.8e3   &     0.1    &   191   & 0.03      & 0.2$\%$     & \nl
      &         &         & \bf{n. 43} &         &           &            & \nl
~~~~~~~~~~~~~~~~~~~~~~~~~~~~~~~~~~~~~~ &  DIM    &   3.1e3   &     0.1    &   221   & 0.02      & 30$\%$      & \nl
~~~~~~~~~~~~~~~~~~~~~~~~~~~~~~~~~~~~~~ &  HII    &   3.1e3   &     0.05   &   181   & 0.03      & 1.7$\%$     & \nl
      &         &         & \bf{n. 45} &         &           &            & \nl
~~~~~~~~~~~~~~~~~~~~~~~~~~~~~~~~~~~~~~ &  DIM    &   3.1e2   &     1.8    &   337   & 0.02      & 33$\%$      & \nl
~~~~~~~~~~~~~~~~~~~~~~~~~~~~~~~~~~~~~~ &  HII    &   1.0e1   &     3.1    &   889   & 0.4       & 1.9$\%$     & \nl
      &         &        & \bf{n. 47}  &         &           &            & \nl
~~~~~~~~~~~~~~~~~~~~~~~~~~~~~~~~~~~~~~ &  DIM    &   1.8e3   &     0.2    &   229   & 0.02      & 21$\%$     & \nl
~~~~~~~~~~~~~~~~~~~~~~~~~~~~~~~~~~~~~~ &  HII    &   1.8e3   &     0.1    &   191   & 0.03      & 1.2$\%$    & \nl
\enddata
\label{tab5}
\end{deluxetable}

\begin{deluxetable}{rrrrrrr}
\small
\tablewidth{3.5in}
\tablenum{6}
\tablecaption{Same as Table 5 but for NGC1313.  }
\tablehead{
\colhead{Type}&\colhead{n}&\colhead{$\frac{G_0}{n}$}&\colhead{T$_s$}&\colhead{$\Phi_A$}&\colhead{IM}&\nl
\colhead{}&\colhead{cm$^{-3}$}&\colhead{}&\colhead{K} & \colhead{} &\colhead{$\%$} 
&\nl
\colhead{(1)}&\colhead{(2)}&\colhead{(3)}& \colhead{(4)} &\colhead{(5)}
&\colhead{(6)}&\nl
}
\startdata 
      &          &    NGC1313    &       &       &             & \nl
\hline
      &          &  \bf{n. 89}   &       &       &             & \nl
DIM   &  1.8e3   &   0.3         & 268   & 0.01  & $< $42$\%$  & \nl
HII   &  1.8e3   &   0.1         & 191   & 0.03  & $< $ 2$\%$  & \nl
      &          &  \bf{n. 90}   &       &       &             & \nl
DIM   &  3.1e3   &   0.2         & 262   & 0.01  & $< $32$\%$  & \nl
HII   &  1.8e3   &   0.2         & 229   & 0.02  & $< $ 2$\%$  & \nl
      &          &   \bf{n. 91}  &       &       &             & \nl
DIM   &  \nodata &   \nodata     &\nodata&\nodata& 100$\%$     & \nl
HII   &  1.8e3   &   0.1         & 191   &0.008  & $<$ 7$\%$   & \nl
\enddata
\normalsize
\label{tab6}
\end{deluxetable}

\begin{deluxetable}{cccc}
\small
\tablecomments{1 K km s$^{-1}$ = 3.39$\times$10$^{-9}$ erg s$^{-1}$ cm$^{-2}$
sr$^{-1}$}
\tablewidth{3.5in}
\tablenum{7}
\tablecaption{ $^{12}$CO(1--0) intensities in the LWS beam of those  regions
inside NGC6946 included in the CO BIMA-SONG map.}
\def\x{$\times$}
\tablehead{ 
\colhead{No.}&\colhead{No~ R.A.~~~~~~Dec.}&\colhead{I($^{12}$CO(1-0))  }&\\
&\colhead{~~J2000}&\colhead{K km s$^{-1}$ }&\\ 
\colhead{(1)}&\colhead{(2)}&\colhead{(3))  }&\\
}
\startdata
17~~~~&   20:34:30.8~ 60:07:08.04 &  3.583     &\\
24~~~~&   20:34:31.6~ 60:08:18.96 &  4.943     &\\
33~~~~&   20:34:32.3~ 60:09:28.80 &  3.835     &\\
41~~~~&   20:34:33.0~ 60:10:39.01 &  3.925     &\\
47~~~~&   20:34:33.8~ 60:11:49.92 &   2.97     &\\
16~~~~&   20:34:40.1~ 60:07:01.92 &   4.63     &\\
23~~~~&   20:34:41.0~ 60:08:12.85 &  9.427     &\\
32~~~~&   20:34:41.8~ 60:09:23.76 &  10.22     &\\
40~~~~&   20:34:42.5~ 60:10:33.60 &  5.508     &\\
46~~~~&   20:34:43.2~ 60:11:43.80 &   2.64     &\\
15~~~~&   20:34:49.6~ 60:06:56.52 &  7.622     &\\
22~~~~&   20:34:50.3~ 60:08:06.72 &  19.07     &\\
31~~~~&   20:34:51.2~ 60:09:17.64 &  30.36     &\\
39~~~~&   20:34:52.0~ 60:10:28.56 &  12.32     &\\
45~~~~&   20:34:52.7~ 60:11:38.40 &  4.032     &\\
14~~~~&   20:34:59.1~ 60:06:51.12 &  5.785     &\\
21~~~~&   20:34:59.9~ 60:08:02.04 &  11.72     &\\
30~~~~&   20:35:00.6~ 60:09:11.16 &   16.2     &\\
38~~~~&   20:35:01.4~ 60:10:22.44 &  9.443     &\\
44~~~~&   20:35:02.2~ 60:11:33.00 &  5.551     &\\
13~~~~&   20:35:08.7~ 60:06:45.00 &  4.227     &\\
20~~~~&   20:35:09.3~ 60:07:55.91 &  6.317     &\\
29~~~~&   20:35:10.1~ 60:09:06.12 &  7.775     &\\
37~~~~&   20:35:10.8~ 60:10:15.96 &  6.526     &\\
43~~~~&   20:35:11.6~ 60:11:27.24 &  3.907     &\\

\enddata
\normalsize
\label{tab7}
\end{deluxetable}

\begin{deluxetable}{cccc}
\small
\tablecomments{ 1 K km s$^{-1}$ = 1.6$\times$10$^{-9}$ erg s$^{-1}$ cm$^{-2}$
sr$^{-1}$} 
\tablewidth{3.5in}
\tablenum{8}
\tablecaption{ $^{12}$CO(1--0) intensities in the LWS beam of three regions in
 NGC1313. CO data are from SEST and they will presented in a  future paper
by Rubio \etal (in preparation).}
\def\x{$\times$}
\tablehead{
\colhead{No.}&\colhead{No~ R.A.~~~~~~Dec.}&\colhead{I($^{12}$CO(1-0))  }&\\
&\colhead{~~J2000}&\colhead{K km s$^{-1}$}&\\ 
\colhead{(1)}&\colhead{(2)}&\colhead{(3))  }&\\
}
\startdata
89~~~~&   3:18:09.0~  -66:28:36.5 &    0.48      &\\
90~~~~&   3:18:09.0~  -66:29:59.6 &    0.58      &\\
69~~~~&   3:18:47.0~  -66:30:00.3 &    0.79      &\\
\enddata
\normalsize
\label{tab8}
\end{deluxetable}

\begin{deluxetable}{cccccc}
\small
\tablewidth{3.5in}
\tablenum{9}
\tablecaption{ LWS beam averaged atomic hydgrogen, 
molecular hydrogen and total hydrogen nucleus column
densities for regions 69, 89 and 90 in NGC1313. N(H$_2)$  has been derived 
from CO intensities assuming a conversion factor
X=1.6$\times$10$^{21}$ cm$^{-2}$ calculated from Eq. 3.2 of Israel 1999 and
taking into account the average metallicity of NGC1313.}
\tablehead{
\colhead{No.}&\colhead{N(HI)}&\colhead{N(H$_2)$
}&\colhead{N(HI)+2$\times$N(H$_2$) }&\\
             &\colhead{10$^{21}$ cm$^{-2}$}&\colhead{10$^{21}$ cm$^{-2}$}&\colhead{10$^{21}$ cm$^{-2}$}&\\ 
\colhead{(1)}&\colhead{(2)}&\colhead{(3))  }& \colhead{(4)  }& 
}
\startdata
69~~~~& 0.9   & 1.2  	  & 3.5  &\\
89~~~~& 1.4   & 0.7 	  & 2.9  &\\
90~~~~& 1.2   & 0.9 	  & 3.0  &\\
\enddata
\normalsize
\label{tab9}
\end{deluxetable}

\begin{deluxetable}{ccccccccccccccccc}
\small
\tablewidth{7.0in}
\tablenum{10}
\tablecaption{Percentage of [CII] emission and HI column density arising from
 dense gas as defined in \S5.4, for   those regions
of NGC6946  observed in the
[CII(158~\micron)],
[OI(63~\micron)] and NII(122~\micron)] lines and corresponding PDR model solutions.
Col. (1): Type of ionized medium (IM) contributing to the [CII] observed flux.
Col. (2) and (3): Percentage of [CII] emission and HI column density arising in dense 
(n$\gg$ 3$\times$10$^3$ cm$^{-3}$) atomic gas. 
Col. (4) and (5): n and $G_0$ model solutions for the dense gas component
only. 
}
\tablehead{
\colhead{~~~~~~~~~~~~~~~~~~~~~~~~~~~~~~~~~~~~~~}&\colhead{Type IM}&\colhead{CII$^{d}$ } &\colhead{HI$^{d}$ }  &\colhead{n$^d$ } 
& \colhead{G$^d_0$ }&\nl
\colhead{~~~~~~~~~~~~~~~~~~~~~~~~~~~~~~~~~~~~~~}&\colhead{}&\colhead{$\%$} &  \colhead{ $\%$}&\colhead{cm$^{-3}$ }  &\colhead{ }
&  \nl
\colhead{~~~~~~~~~~~~~~~~~~~~~~~~~~~~~~~~~~~~~~}&\colhead{(1)}&\colhead{(2)}&\colhead{(3)}& \colhead{(4)} &\colhead{(5)} &\nl 
}
\startdata
                   &            &   NGC6946    &        &               &\nl
\hline
                   &            & \bf{n. 15}   &        &               &\nl
~~~~~~~~~~~~~~~~~~~~~~~~~~~~~~~~~~~~~~ & DIM   &   74$\%$   & 3.7$\%$      & 1.8e3  & 5.6e2         &\nl
~~~~~~~~~~~~~~~~~~~~~~~~~~~~~~~~~~~~~~ & HII   &   81$\%$   & 6.0$\%$      & 1.8e3  & 3.1e2         &\nl
                   &            & \bf{n. 21}   &        &               &\nl
~~~~~~~~~~~~~~~~~~~~~~~~~~~~~~~~~~~~~~ & DIM   & $< $78$\%$ & $< $5.0$\%$  & 1.8e3  & 1.0e3         &\nl
~~~~~~~~~~~~~~~~~~~~~~~~~~~~~~~~~~~~~~ & HII   & $< $87$\%$ & $< $9.0$\%$  & 1.0e3  & 3.1e2         &\nl
                   &            & \bf{n. 23}   &        &               &\nl
~~~~~~~~~~~~~~~~~~~~~~~~~~~~~~~~~~~~~~ & DIM   &    72$\%$  &  4.0$\%$     & 3.1e3  &  1.8e3        &\nl
~~~~~~~~~~~~~~~~~~~~~~~~~~~~~~~~~~~~~~ & HII   &    85$\%$  &  7.8$\%$     & 1.8e3  &  5.6e2        &\nl
                   &            & \bf{n. 29}   &        &               &\nl
~~~~~~~~~~~~~~~~~~~~~~~~~~~~~~~~~~~~~~ & DIM   &    80$\%$  &  5.6$\%$     & 1.8e3  & 1.8e3         &\nl
~~~~~~~~~~~~~~~~~~~~~~~~~~~~~~~~~~~~~~ & HII   &    89$\%$  &  9.4$\%$     & 5.6e2  & 5.6e2         &\nl
                   &            & \bf{n. 31}   &        &               &\nl
~~~~~~~~~~~~~~~~~~~~~~~~~~~~~~~~~~~~~~ & DIM   &   92$\%$   & 14$\%$       & 1.8e3  &  1.0e3        &\nl
~~~~~~~~~~~~~~~~~~~~~~~~~~~~~~~~~~~~~~ & HII   &   95$\%$   & 25$\%$       & 1.0e3  &  3.1e2        &\nl
                   &            & \bf{n. 33}   &        &               &\nl
~~~~~~~~~~~~~~~~~~~~~~~~~~~~~~~~~~~~~~ & DIM   &   58$\%$   & 2.6$\%$      & 5.6e3  &   1.8e3       &\nl
~~~~~~~~~~~~~~~~~~~~~~~~~~~~~~~~~~~~~~ & HII   &   75$\%$   & 4.4$\%$      & 1.8e3  &   5.6e2       &\nl
                   &            & \bf{n. 38}   &        &               &\nl
~~~~~~~~~~~~~~~~~~~~~~~~~~~~~~~~~~~~~~ & DIM   &   73$\%$   & 4.0$\%$      & 3.1e3  &  3.1e3        &\nl
~~~~~~~~~~~~~~~~~~~~~~~~~~~~~~~~~~~~~~ & HII   &   89$\%$   & 10$\%$       & 1.0e3  &  5.6e2        &\nl
                   &            & \bf{n. 40}   &        &               &\nl
~~~~~~~~~~~~~~~~~~~~~~~~~~~~~~~~~~~~~~ & DIM   &   70$\%$   & 3.7$\%$      & 3.1e3  &  5.6e2        &\nl
~~~~~~~~~~~~~~~~~~~~~~~~~~~~~~~~~~~~~~ & HII   &   79$\%$   & 5.4$\%$      & 3.1e3  &  3.1e2        &\nl
                   &            & \bf{n. 43}   &        &               &\nl
~~~~~~~~~~~~~~~~~~~~~~~~~~~~~~~~~~~~~~ & DIM   &   75$\%$   & 4.4$\%$      & 3.1e3  &  3.1e2        &\nl
~~~~~~~~~~~~~~~~~~~~~~~~~~~~~~~~~~~~~~ & HII   &   82$\%$   & 6.2$\%$      & 3.1e3  &  1.8e2        &\nl
                   &            & \bf{n. 45}   &        &               &\nl
~~~~~~~~~~~~~~~~~~~~~~~~~~~~~~~~~~~~~~ & DIM   &   61$\%$   & 2.8$\%$      & 1.0e3  &   1.8e3       &\nl
~~~~~~~~~~~~~~~~~~~~~~~~~~~~~~~~~~~~~~ & HII   &   73$\%$   & 4.2$\%$      & 3.1e2  &   5.6e2       &\nl
                   &            & \bf{n. 47}   &        &               &\nl
~~~~~~~~~~~~~~~~~~~~~~~~~~~~~~~~~~~~~~ & DIM   &   71$\%$   & 3.9$\%$      & 1.8e3  &   5.6e2       &\nl
~~~~~~~~~~~~~~~~~~~~~~~~~~~~~~~~~~~~~~ & HII   &   78$\%$   & 4.8$\%$      & 1.8e3  &   3.1e2       &\nl
\enddata
\label{tab10}
\end{deluxetable}

\begin{deluxetable}{rrrrrrr }
\small
\tablewidth{3.5in}
\tablenum{11}
\tablecaption{Same as Table 10 but for NGC1313.  }
\tablehead{
\colhead{Type}&\colhead{CII$^d$ }  &\colhead{HI$^{d}$ } & \colhead{n$^d$ }& 
\colhead{G$^d_0$ } & \nl
\colhead{}&\colhead{$\%$ }  &\colhead{$\%$ }  &\colhead{cm$^{-3}$}&\colhead{}&\nl
\colhead{(1)}&\colhead{(2)}&\colhead{(3)}& \colhead{(4)} &\colhead{(5)}&\nl
}
\startdata 
        &           &  NGC1313    & 	        &	   & \nl
\hline
      &             &  \bf{n. 89} &             &         &  \nl
DIM   &  $< $36$\%$ & $< $1.7$\%$ & 1.8e3       &  5.6e3  &  \nl
HII   &  $< $62$\%$ & $< $3.0$\%$ & 1.8e3       &  3.1e2  &  \nl
      &             &  \bf{n. 90} &             &         &  \nl
DIM   &  $< $45$\%$ & $< $2.0$\%$ & 1.8e3       &  5.6e3  &  \nl
HII   &  $< $62$\%$ & $< $3.0$\%$ & 5.6e2       &  3.1e3  &  \nl
      &             &  \bf{n. 91} &             &         &  \nl
DIM   & \nodata     & \nodata     & \nodata     & \nodata &  \nl
HII   & $< $4.5$\%$ &$<$1.0$\%$   & 3.1e5       &  1.0e1  &  \nl
\enddata
\normalsize
\label{tab11}
\end{deluxetable}
 
\newpage

\begin{figure}
\caption{
The LWS pointings at different wavelengths on  HI maps for NGC1313 
(Ryder \etal  1995, resoltuion equal to 29\arcsec$\times$28\arcsec; top panel) 
and NGC6946 (Boulanger $\&$ Viallefond 1992, resolution equal to
 24.5\arcsec$\times$28.3\arcsec ; bottom panel). 
LWS pointings are represented by circles of diameter equal
to the FHWM of the LWS beam at 158 $\mu$m (75$\arcsec$). Black ellipses represent the
optical sizes of the galaxies at surface brightness equal to 25$^{th}$ magnitude in the B
band (R$_{25}$).
Top panel: NGC1313. All fields indicated by circles were observed at 158 $\mu$m. Blue circles 
correspond to pointings observed at 63 $\mu$m  and 122 $\mu$m.
Bottom panel: NGC6946. All circles were observed at 158
$\mu$m. Blue, green and yellow circles correspond to the pointings observed 
at 63 $\mu$m; green and 
yellow circles  at 122 $\mu$m, yellow circles at 88 $\mu$m.
}
\end{figure}  

\begin{figure}
\caption{
HI contours superposed on an IRAS 
map at 100 $\mu$m of NGC1313 (grey scale). The position of the point observed by
 LWS to measure the [CII] foreground emission is marked (REF). Note the elevated levels of
Galactic cirrus
in the direction of NGC1313 and the lower foreground emission around the reference
 position observed with LWS.
}
\end{figure}

\begin{figure}
\caption{
The NGC1313 (top) and NGC6946 (bottom) [CII] contours superposed on 
the LW2 (6.75 $\mu$m) ISOCAM images (Dale et al. 2000).
Only the [CII] flux values higher than 5 $\sigma$, after background
subtraction, 
have been considered for the
interpolation.
The contour levels for NGC6946 go from 1.6 to 
20$\times$10$^{-6}$ erg s$^{-1}$ cm$^{-2}$sr$^{-1}$
with 0.8$\times$10$^{-6}$ erg s$^{-1}$ cm$^{-2}$sr$^{-1}$ spacing. 
For NGC1313 they go from 1.8 to 
9.5$\times$10$^{-6}$ erg s$^{-1}$ cm$^{-2}$sr$^{-1}$ with
0.4$\times$10$^{-6}$ erg s$^{-1}$ cm$^{-2}$sr$^{-1}$ step.
}
\end{figure}

\begin{figure}
\caption{
Top panel: log ([CII]/FIR) as a function of  60/100 $\mu$m
ratio which is indicative of the radiation field intensity. The 5$\sigma$ detections
at 158 $\mu$m inside NGC1313 and NGC6946 are
plotted together with the global ratios of a sample of 60 normal galaxies
(Malhotra et al. 2000, 2001). The Irregular galaxies belonging to the ISO--KP
sample are marked with large squares and three regions inside IC10 with
asterisks (Hunter et al. 2001). 
Bottom panel: same as the top panel but for log([CII]/$\nu$ f$_{\nu}$(5-10 $\mu$m)).
The $\nu$ f$_{\nu}$(5-10$\mu$m) represents the contribution of the Aromatic 
Feature Emission between 5 and 10 $\mu$m. This quantity has been derived from the ISOCAM
LW2 (5--8.5 $\mu$m) observations as in Helou et al. (2001). 
For display purpose we do not plot the single error measurements but
the average errors are marked in the legend symbols.
The upper
limits with the highest 60/100 $\mu$m ratio  among the whole sample of 60
normal galaxies represents NGC4418.   As explained in Helou \etal (2000) and Lu
et al. (2000),
the MIR spectrum of this galaxy differs from the typical spectra of normal
galaxies, indicating that for this galaxy the flux between 5 and 10 $\mu$m does
not arise from the classical carriers producing AFE. See also 
Spoon et al. (2001).
}
\end{figure}

\begin{figure}
\caption{ 
Top left panel: the histogram of the log ([CII]/FIR) ratio for NGC6946 and
NGC1313.  
Top right panel:the histogram of the log ([CII]/$\nu$ f$_{\nu}$(5-10 $\mu$m)) ratio for NGC6946 and
NGC1313.  
Middle left panel: the logarithm of the FIR flux (per beam) distribution. 
Middle right panel: the  logarithm of the $\nu$ f(5-10$\mu$m) flux  (per beam) distribution.  
Bottom left panel: the  logarithm of the [CII] flux  (per beam) distribution. 
Bottom right panel: the histogram of the $\nu$ f(5-10$\mu$m)/FIR ratios. 
The symbol $m$ ands $\sigma$ indicate the mean value of the distribution and its
$\sigma$ respectively.
}
\end{figure}

\begin{figure}
\caption{
Top. North--East South--West cut of NGC6946 with P.A. =45$\degr$ along
 the LWS pointings. For each wavelength, the intensities are normalized to
the maximum value. Bottom: same as the top panel but with a P.A. = 135$\degr$.
}
\end{figure}

\begin{figure}
\caption{
Results from the PDR model by Kaufman et al. 1999 for the 
regions 
observed with LWS in  NGC6946 (triangles) and
NGC1313 (circles). Only those regions with [OI(63 $\mu$m)], [NII(122 $\mu$m)] and
[CII(158 $\mu$m)] have been used.   Also plotted, the results obtained from the integrated emission of the
ISO--KP sample by Malhotra \etal 2001 (stars) and the values of G$_0$
and n obtained averaging the results of the  regions considered in NGC6946 
(filled triangle). 
}
\end{figure}

\begin{figure}
\caption{
Regions in NGC6946  observed in the [CII(158 $\mu$m)], OI(63 $\mu$m)]
and NII(122 $\mu$m)] fine structure lines with LWS, for which it has been possible to apply the model
predictions of Kaufman et al. (1999). The grey scale  is the  ISOCAM LW2 (5--8.5
$\mu$m) image (Dale et al. 2000). The MIR surface brightness is higher in dense and warm
PDRs. 
}
\end{figure}

\begin{figure}
\caption{
[CII(158 $\mu$m)] intensities as a function of the $^{12}$CO(1--0)
intensities for regions in
NGC6946 (triangles), NGC1313 (open circles), galaxies of the ISO--KP
for which CO data are available (Lord et al. in preparation, stars)
and KAO data at 158 $\mu$m for the Stacey et al. (1991) sample (squares).
Filled squares represent the warmer galaxies of the Stacey sample.
The dashed line represents the relation followed by warm galaxies and Galactic
star forming regions with a mean I$_{[CII]}$/I$_{CO}$ ratio of about 4200.
The solid line represents the relation followed by cooler normal spirals, with
mean ratio 1300.
}
\end{figure}

\begin{figure}
\caption{
The deprojected [CII] surface brightness vs. the deprojected HI column
 density  for NGC1313 (circles) and NGC6946 (triangles). 
Also plotted as squares is the sample
from Stacey et al. (1991). Within the Stacey sample is the integrated emission 
of the nucleus (55$\arcsec$) of NGC6946, marked on the figure with an asterisk. The dashed horizontal lines  
correspond to  the Nucleus (N), Spiral Arms (SA) and Extended (E) emission component from Madden \etal
(1993). The solid lines correspond to the calculated [CII] emission from 
 standard HI clouds (n$\sim$ 30 and
T$\sim$ 100, Eq. A2 from Crawford \etal 1985
1985) and that of the intercloud medium (n$\sim$ 0.1 and T$\sim$10$^4$ Eq. A3 of
Crawford \etal 1985) assuming area filling factors equal to unity.
The dashed diagonal line represents the  observed galactic cirrus emission
(Bennett et al. 1994).
}
\end{figure}

\begin{figure}
\caption{ 
The CII [(158 $\mu$m)]/[NII[122 $\mu$m)] ratio as a function
of the [CII]/HI ratio for those regions observed at 122 $\mu$m in
NGC6946 (triangles) and NGC1313 (circles).
}
\end{figure}

\end{document}